\documentclass[11pt]{article}
\usepackage{epsfig}
\usepackage{latexsym}

\begin{document}
\thispagestyle{empty}
\begin{flushright}
hep-lat/0212029
\end{flushright}
\begin{center}
\vspace*{6mm}
{\LARGE New findings for topological excitations\\
\vspace{2mm}
in SU(3) lattice gauge theory}
\vskip10mm
{\bf Christof Gattringer and Stefan Schaefer}
\vskip3mm
Institut f\"ur Theoretische Physik\\
Universit\"at Regensburg \\
D-93040 Regensburg, Germany
\vskip25mm
\begin{abstract}
We probe the SU(3) vacuum using eigenvectors of the Dirac operator
with an arbitrary phase for the temporal boundary condition. We
consider configurations with topological charge $|Q|$ = 1 
near the QCD phase transition and at low temperatures on a torus.
For all our ensembles we show that 
the zero-mode of the Dirac operator changes its position as one
changes the phase of the boundary condition. For ensembles
near the QCD phase transition our results closely
resemble the behavior of zero-modes for Kraan - van Baal solutions
of the classical Yang-Mills equations where the individual lumps
are interpreted as monopoles. Our findings near $T_c$ and on the torus 
show that for both cases  
an excitation with topological charge $|Q|$ = 1 is built from
several separate lumps.
\end{abstract}
\vspace{4mm}
{\it To appear in Nuclear Physics B.}
\end{center}
\vskip12mm
\noindent
{\sl Due to size limitations of the electronic archive this version 
contains less plots than the complete paper submitted to the journal.  
The long version can be obtained from the authors at:} \\
{\tt christof.gattringer@physik.uni-regensburg.de \\
stefan.schaefer@physik.uni-regensburg.de}

\newpage
\section{Introduction}

QCD exhibits two remarkable features, confinement and chiral symmetry 
breaking, which both are cornerstones for our understanding of strongly 
interacting particles. It is an interesting property
of QCD that at the QCD phase transition the restoration of chiral 
symmetry and the disappearance of confinement occur at the same 
critical temperature. This seems to indicate that the two phenomena
are intimately linked with each other. So far the two phenomena have resisted 
our attempts to find a description unifying them and the structure of 
the fundamental excitations of the QCD vacuum is still unclear. 

In recent years lattice calculations have started to contribute 
to our understanding of the QCD vacuum fluctuations. 
In particular it was found that the eigenmodes of the Dirac operator provide 
a powerful filter removing the hard UV fluctuations and only
the long range
modes couple to the low eigenvectors \cite{negele}-\cite{horvath2}. 
These studies were inspired 
by the phenomenological picture of chiral symmetry breaking based on 
instantons \cite{instantons}. In this picture the QCD vacuum is described as 
a fluid of interacting instantons and anti-instantons. Instead of a 
zero-mode which exists for a single instanton \cite{index,thooftzm} the fluid
of interacting instantons and anti-instantons gives rise to many 
near zero-modes which have small but non-zero eigenvalues. The density
of these eigenvalues is then related to the chiral condensate 
by the Banks-Casher formula \cite{bankscasher}. 
Lattice studies have established that the near zero-modes do indeed show
a lumpy structure and are locally chiral as expected from the instanton 
picture. Furthermore it was demonstrated that the lumps in the field strength
have a high degree of self-duality \cite{selfdual}. Note however, that first
attempts to fit eigenmodes of the Dirac operator with the profile
of the 't Hooft zero-mode have failed \cite{horvath2}. 

In this article we use a new method for probing the QCD vacuum with 
eigenvectors of the lattice Dirac operator. We introduce an arbitrary 
phase $\exp(i 2\pi \zeta)$ with $\zeta \in [0,1]$ at the temporal boundary 
of the Dirac operator and analyze how the zero-mode in configurations 
with topological charge $Q = \pm 1$ responds to changes in $\zeta$. 
The most prominent feature we discover is that the zero-mode changes its 
position and can be located at different space-time points for different
values of $\zeta$. This property is observed for quenched ensembles 
of SU(3) gauge configurations with 
temperature, both in the confined and the deconfined phase as well as for
configurations on a torus.  

Our generalized boundary condition is motivated by an 
interesting property of zero-modes for Kraan-van Baal (KvB)
solutions of the classical Yang-Mills equations on a euclidean cylinder.
KvB solutions \cite{kvb1} generalize the caloron solution 
\cite{HaSh} by allowing for non-trivial Polyakov loop at spatial infinity.
KvB solutions depend in addition to the phases of the 
Polyakov loop at spatial infinity also on $N$ (for SU($N$)) vectors
$\vec{y}_i, i = 1,2 \, ... \, N$. When one draws apart those vectors
one finds that an object of charge 1 is built from $N$ constituent monopoles
and this property might lead to the missing link 
between confinement and chiral symmetry breaking. 
Strong evidence for SU(2) KvB solutions in cooled lattice 
gauge configurations were given 
for twisted \cite{garciaetal99} and periodic boundary 
conditions \cite{kvblat2}.

For KvB solutions also the zero-mode has been computed \cite{kvbzm}. 
This zero-mode has 
the remarkable property that it is located on only one of the constituent
monopoles but it can jump from one monopole to another when changing 
$\zeta$. 
In a previous article \cite{gattringer02} we have compared zero-modes
with periodic 
and anti-periodic boundary conditions and have found clear signals 
for KvB-type behavior. Here we now use the generalized boundary 
condition with phase $\exp(i 2\pi \zeta)$ and present further strong
evidence that the excitations of QCD at high temperature have the structure 
of KvB solutions. 

In a subsequent step we apply the techniques that were successful in 
detecting the constituent monopoles at high temperature to gauge configurations
generated on a torus. For the torus no equivalent of KvB solutions is known.
However, several articles have put forward the idea that also at 
low temperatures constituents with fractional 
charge build up lumps with integer topological charge \cite{arroyo}. 
Our finding, that when changing the fermionic boundary condition the
zero-mode does change its position, indicates that 
indeed an excitation with topological charge $|Q|$ = 1 is built from
several separate lumps.

\section{Technicalities} 

\subsection{Gauge configurations}

For our runs we use gauge ensembles generated with the L\"uscher-Weisz
action \cite{luweact} with coefficients from tadpole improved perturbation
theory. We work at two different values of the inverse gauge coupling
$\beta = 8.20$ and $\beta = 8.45$. A determination of the lattice spacing
based on the Sommer parameter gives $a = 0.115(1)$ fm and $a = 0.094(1)$
respectively \cite{sommerci}. 

At those two couplings we generate configurations on $6\times20^3$ lattices.
The two values of $\beta$ then give rise to an ensemble in the confining
phase ($\beta = 8.20$) and an ensemble in the deconfined phase 
($\beta = 8.45$). The corresponding temperatures are 287 and 350 MeV.
The critical temperature for the QCD phase transition
was determined as $T_c = 300$ MeV for the L\"uscher-Weisz action 
\cite{crittemp}. 

We restrict ourselves to configurations with topological charge
$Q = \pm 1$, by considering configurations with a single zero-mode. 
Thus problems with mixing of different zero-modes are avoided. 
From two larger samples of typically
400 configurations we selected 70 configurations with charge $Q = \pm1$ 
for $\beta = 8.20$ and 89 configurations for $\beta = 8.45$. 

For the deconfined ensemble ($\beta = 8.45$) the Polyakov loop has a 
non-vanishing expectation value. Due to the $Z_3$ symmetry of the gauge 
action the Polyakov loop can come with three different values for its 
phase $\varphi = 0, \pm 2\pi/3$. The Dirac operator does not have this 
symmetry and thus the eigenvectors and eigenvalues of the Dirac operator
will behave differently for real, respectively complex Polyakov loop. 
We found that of the 89 configurations in the deconfined phase 25 have 
real Polyakov loop and 64 have $\varphi = \pm 2\pi/3$. In the following
we will often refer to these two subsets as the real and complex sector. 

Finally we also generated a low temperature ensemble on $16^4$ lattices at 
$\beta = 8.45$. It consists of 46 configurations. For all ensembles we 
use periodic
boundary conditions for the gauge fields. In Table~\ref{confdata}
we summarize the parameters of our ensembles. We quote the lattice size, 
the inverse gauge coupling, the statistics, the lattice spacing and the 
temperature in MeV as well as in units of $T_c$. We formally asign also a
temperature to the $16^4$ ensemble even though no temporal direction is 
distinguished by the lattice geometry. 

\begin{table}[t]
\vspace{-5mm}
\begin{center}
\begin{tabular}{c|c|c|c|c|c}
size & $\beta$ & statistics & $a$ [fm] & $T$ [MeV] & $T/T_c$ \\
\hline
$6\times20^3$ & 8.45 & 64 + 25 & 0.094(1) & 350 & 1.17 \\
$6\times20^3$ & 8.20 & 70      & 0.115(1) & 287 & 0.96 \\
$16^4$        & 8.45 & 46      & 0.094(1) & 113 & 0.44 \\
\end{tabular}
\end{center}
\caption{Parameters of our gauge ensembles.
\label{confdata}}
\end{table}

\subsection{Observables}

As already outlined in the introduction we use the eigenvectors of the Dirac
operators to analyze excitations of the QCD vacuum since they provide an 
excellent filter removing the UV fluctuations. Here we use the chirally
improved operator \cite{ci}. The chirally improved Dirac operator is a 
systematic expansion of a solution of the Ginsparg-Wilson equation 
\cite{GiWi82} and has very good chiral properties. It has been successfully
used for quenched spectroscopy \cite{boston}
with pion masses down to 230 MeV. In 
\cite{insttest} it was shown, that chirally improved 
fermions reproduce the analytic result 
for the zero-mode in the background of a discretized instanton very well.
In particular for small instanton radii it is considerably more accurate
than the overlap operator. 

We compute the eigenvectors and eigenvalues of the chirally improved 
Dirac operator using the implicitly restarted Arnoldi method \cite{arnoldi}. 
For the fermions we use periodic boundary conditions for the 
spatial directions. In time direction we implement general
boundary conditions that allow for an arbitray phase,
\begin{equation}
\psi(N_t + 1, \vec{x}) \; \; = \; \; \exp(\,i \,2\pi \, \zeta) \, 
\psi(1, \vec{x}) \; .
\label{zetabc}
\end{equation}
By $N_t$ we denote the number of lattice points in time direction. The
parameter $\zeta$ can assume values in the interval [0,1], 
with $\zeta = 0$ corresponding
to periodic temporal boundary conditions and $\zeta = 0.5$ giving
the anti-periodic case. Note that the value $\zeta = 1$ gives the same
boundary condition as $\zeta = 0$. 
For all of our configurations as listed 
in Table~\ref{confdata} we computed eigenmodes for periodic and
anti-periodic temporal boundary conditions. For smaller subsets
from each ensemble we also computed eigenvectors for smaller steps
in $\zeta$, such as $\zeta = 0.1, 0.2 \, ...$ . More detailed information
on these configurations is listed in tables below.

From the eigenvectors $\psi$ we construct several gauge invariant observables.
The simplest observable is the scalar density $\rho(x)$ of the eigenvectors
obtained by summing color and Dirac indices $c,d$ for each space-time point $x$
separately,
\begin{equation}
\rho(x) \; \; = \; \; \sum_{c,d} |\psi(x)_{c,d}|^2 \; .
\label{rhodef}
\end{equation}
Note that since our eigenvectors $\psi$ are normalized one has
$\sum_x \rho(x) = 1$ where the sum runs over all lattice points $x$.
From the analytic continuum result as well as from lattice studies
it is known that the zero-mode 
traces the underlying instanton. 
Plots of the density $\rho(x)$ for slices of the lattice will be used 
below to illustrate the behavior of the eigenmodes. We will also
compare the position of the lumps seen by different 
eigenmodes $\psi_1,\psi_2$
by computing the euclidean 4-distance $d_4$ between their maxima
\begin{equation}
d_4 \; \; = \; \; \parallel x^{max}_1 - x^{max}_2 \parallel \; ,
\end{equation}
where $x^{max}_1$ and $x^{max}_2$ are the lattice points with 
the largest values of the density for $\psi_1$, respectively $\psi_2$.

In order to quantify the localization of the lump seen by the eigenmode
we use the so-called inverse participation ratio defined as
\begin{equation}
I \; \; = \; \; V \sum_x \rho(x)^2 \; ,
\label{iprdef}
\end{equation}
where $V$ denotes the total number of lattice points. If a mode
is localized on a single site $x_0$ ($\rho(x) = \delta_{x,x_0}$) then
one finds $I = V$. For an entirely spread out mode ($\rho(x) = 1/V$ for 
all $x$) one has $I=1$. Thus localized modes have large $I$, while 
delocalized modes have small values of $I$.

Finally we describe an observable to characterize the overlap of lumps
seen by different eigenmodes. In particular we compute the overlap of the
support of the two lumps seen by eigenvectors
$\psi_1$ and $\psi_2$. The support is defined as the function $\theta(x)$
which has $\theta(x) = 1$ for the lattice points $x$ carrying the 
lump in $\rho$ and $\theta(x) = 0$ for all other points. 
An open question is how to identify the
lump. For a narrow lump the support is small while a broad lump
has a much larger support. 
Using the inverse participation ratio 
$I$ we define a number $N =$ Int$[V/(32 I) + 1]$ and use this to 
determine the support by setting $\theta(x) = 1$ for those $N$ lattice 
points $x$ which carry the $N$ largest values of $\rho$ and $\theta(x) = 0$ 
for all other points. The normalization of $N$ is chosen such that if the 
lump was a 4-D Gaussian the support would consist of the 4-volume 
carrying the points inside the half-width of the Gaussian. Note that
$N$ is the size of the support, i.e.~$N = \sum_x \theta(x)$.

With our definition of the support of
the two lumps ($\theta_1$ for eigenvector $\psi_1$, 
$\theta_2$ for eigenvector $\psi_2$) we can define the 
overlap function $C$ of the two lumps as 
\begin{equation}
C(x) \; \; = \; \; \theta_1(x)\theta_2(x) \; .
\end{equation}
From $C(x)$ we compute the relative overlap 
\begin{equation}
R_{12} \; \; = \; \; \frac{2}{N_1 \, + \, N_2} \; \sum_x C(x) \; .
\end{equation}
The relative overlap $R_{12}$ is a real number ranging from 0 to 1. If the 
lump in $\rho_1$ and the lump in $\rho_2$ sit on top of each other 
(at least the core up to about the half-width) and have the same size 
then $R_{12} = 1$. If the two lumps are entirely separated one has 
$R_{12} = 0$. Values in between give the ratio of the common volume of the 
two supports to the average size of the 
two supports. Note that even if the two lumps sit on top of each other,
but have different sizes $N_1, N_2$ one has $R_{12} < 1$.

\section{A few words on Kraan - van Baal solutions}

When we present our data in the sections below we will show that 
the two ensembles with finite temperature, i.e.~the runs on the 
$6\times 20^3$ lattices, can be described by Kraan - van Baal
(KvB) solutions. In order to facilitate this enterprise let us briefly 
discuss the structure of the KvB solutions \cite{kvb1} 
and the analytic results for the corresponding zero-mode \cite{kvbzm}. 

\subsection{Kraan - van Baal solutions}

Kraan van-Baal solutions are classical solutions of the Yang-Mills equations 
on a euclidean cylinder. They are characterized by the Polyakov loop 
at spatial infinity ${\cal P}_\infty$ given by 
\begin{equation}
{\cal P}_\infty \; \; = \; \; \lim_{|\vec{x}| \rightarrow \infty} 
P \, \exp\left( \int_0^\beta dt A_0(t,\vec{x}) \right) \; ,
\end{equation}
where $\beta$ is the time extent of the cylinder and the temporal 
boundary conditions of the gauge field
are periodic, i.e.~$A_\mu(\beta + t,\vec{x}) = A_\mu(t,\vec{x})$. 
With a suitable constant gauge transformation ${\cal P}_\infty$
can be diagonalized and is given by (we show the formulae for SU(3))
\begin{equation}
{\cal P}_\infty \; \; = \; \; \exp\Big( \, i \, 2 \pi \, \mbox{diag} (
\mu_1, \mu_2, \mu_3 ) \Big) \; .
\end{equation}
The determinant of ${\cal P}_\infty$ has to equal 1 which implies that
$\mu_1 + \mu_2 + \mu_3$ is an integer. In order to have a unique 
labeling of ${\cal P}_\infty$ one chooses the constant gauge transformation
and the phases such that 
\begin{eqnarray}
&&\mu_1 \, + \, \mu_2 \, + \, \mu_3 \; \; = \; \; 0 \; ,
\nonumber \\
&&\mu_1 \; \leq \; \mu_2 \; \leq \; \mu_3 \; \leq \; \mu_4 \; \;  
\equiv \; \; 1 + \mu_1 \; .
\end{eqnarray}
In the second line we defined another phase factor $\mu_4$ which will be
useful for a compact notation. Using the scale invariance of the classical 
equations the time extent $\beta$ can be set to $\beta = 1$. Subsequently
we use this convention and distances are now expressed in units of $\beta$.

In addition to the phase factors $\mu_i$ the KvB solution 
also depends on 3 spatial vectors $\vec{y}_1, \vec{y}_2, \vec{y}_3$.
As already remarked the KvB solution can be seen to consist of $3$
constituent monopoles and the $\vec{y}_i$ are their positions.
As for the $\mu_m$ we define a 4-th vector $\vec{y}_4 = \vec{y}_1$
for notational convenience.
The action density of the KvB solutions can be written as
\begin{eqnarray}
\mbox{Tr} F_{\mu\nu}^{\,2}(x) & = & \partial_\mu^2\partial_\nu^2\log\psi(x) \;,
\nonumber \\
\psi(x) & = & \frac{1}{2} \, \mbox{tr}({\cal A}_3 \, 
{\cal A}_2 \, {\cal A}_1 ) 
\; - \; \cos(2\pi t) \;,
\nonumber \\
{\cal A}_m  & = & \frac{1}{r_m}
\left( \begin{array} {cc}
r_m    &   |\vec{y}_m - \vec{y}_{m+1}|  \\ 
0      &   r_{m+1}                    
\end{array} \right) \left(
\begin{array}{cc}
c_m & s_m \\
s_m & c_m
\end{array} 
\right) \; .
\end{eqnarray}
In these equations $r_m = | \vec{x} - \vec{y}_m |$,
$c_m = \cosh(2\pi \nu_m r_m)$ and $s_m = \sinh(2\pi \nu_m r_m)$.
Each monopole can be assigned a mass $M_m$, 
\begin{equation}
M_m \; \; = \; \; 8 \pi^2 \, ( \mu_{m+1} - \mu_m ) \; . 
\label{monomass}
\end{equation}
This expression for the monopole mass $M_m$ implies that the $m^{th}$
monopole is localized when $\mu_{m+1} - \mu_m$ is large and it is 
spread out for small $\mu_{m+1} - \mu_m$. 
  
\begin{figure}[t]
\vspace{5mm}
\begin{center}
\epsfig{file=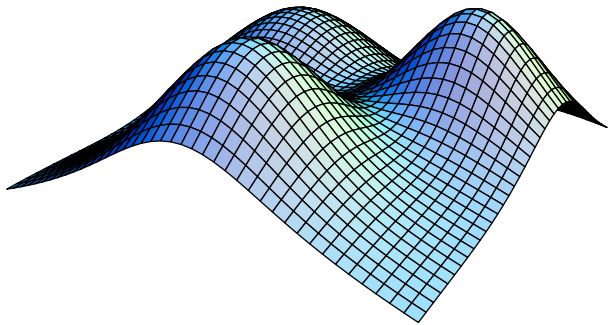,height=5cm,clip}
\end{center}
\vspace{-3mm}
\caption{Action density of a typical KvB solution for a $x,y$-slice
on a logarithmic scale. 
The parameters of the solution are $(\mu_1,\mu_2,\mu_3) =
(-17,-2,19)/60$, $\vec{y}_1 = (-2,-2,0)$, $\vec{y}_2 = (0,2,0)$ and 
$\vec{y}_3 = (2,-1,0)$. 
\label{f2example}}
\end{figure}

In order to illustrate the structure of the KvB solutions we show the
corresponding action density for an example 
in Fig.~\ref{f2example}. The values of the 
parameters\footnote{These values are directly taken from the example given in 
the second reference in \cite{kvbzm}.} are $(\mu_1,\mu_2,\mu_3) =
(-17,-2,19)/60$. The positions of the monopoles are chosen as
$\vec{y}_1 = (-2,-2,0)$, $\vec{y}_2 = (0,2,0)$ and 
$\vec{y}_3 = (2,-1,0)$. We show the action density in the $x,y$-plane
on a logarithmic scale. 

\subsection{Zero-modes of the Dirac operator}

The eigenvalue problem of the Dirac operator in the background of a KvB 
solution was addressed in \cite{kvbzm} and an explicit expression for
the zero-mode was given. In particular the problem was solved for general
temporal boundary conditions (note that we scaled the temporal 
extent $\beta$ to $\beta = 1$),
\begin{equation}
\psi(1 + t, \vec{x}) \; \; = \; \; \exp(\,i \,2\pi \, \zeta) \, 
\psi(t, \vec{x}) \; .
\end{equation}
The density of the zero-mode is given by
\begin{equation}
\rho(x) \; \; = \; \; \sum_{c,d} |\psi(x)_{c,d}|^2 \; \; = \; \; 
- \, \frac{1}{(2\pi)^2} \, \partial_\mu^2 \hat{f}_{\zeta} (x) \; ,
\label{zm1}
\end{equation}
with
\begin{equation}
\hat{f}_{\zeta} (x) \; \; = \; \; \frac{\pi}{r_m} \, \psi(x)^{-1} \,
\langle v^m_\zeta | {\cal A}_{m-1} \, ... \, {\cal A}_1 
{\cal A}_3 \, ... \, {\cal A}_{m}
| w^m_\zeta \rangle \; ,
\label{zm2}
\end{equation}
and the 2-spinors $| v^m_\zeta \rangle$ and $| w^m_\zeta \rangle$
are given by
\begin{equation}
| v^m_\zeta \rangle \; = \, \left(\!\! 
\begin{array} {c}
\sinh(2\pi(\zeta \! - \! \mu_m) r_m) \\
\cosh(2\pi(\zeta \! - \! \mu_m) r_m) 
\end{array} \!\!\right) , \; 
| w^m_\zeta \rangle \; = \, \left(\!\! 
\begin{array} {c}
\cosh(2\pi(\zeta \! - \! \mu_m) r_m) \\
- \sinh(2\pi(\zeta \! - \! \mu_m) r_m) 
\end{array} \!\! \right).
\label{zm3}
\end{equation}
The index $m$ which determines the prefactor $1/r_m$,
the spinors $| v^m_\zeta \rangle$, $| w^m_\zeta \rangle$ 
and the ordering of the ${\cal A}_i$ in 
Eq.~(\ref{zm2}) as well as the choice of $\mu_m$ and $r_m$ in 
Eq.~(\ref{zm3}) is determined from the 
interval containing the boundary condition parameter, 
\begin{equation}
\mu_m \; \leq \; \zeta \; \leq \; \mu_{m+1} \; . 
\label{zmloc}
\end{equation}

\begin{figure}[t!]
\begin{center}
\hspace*{-5mm}
\epsfig{file=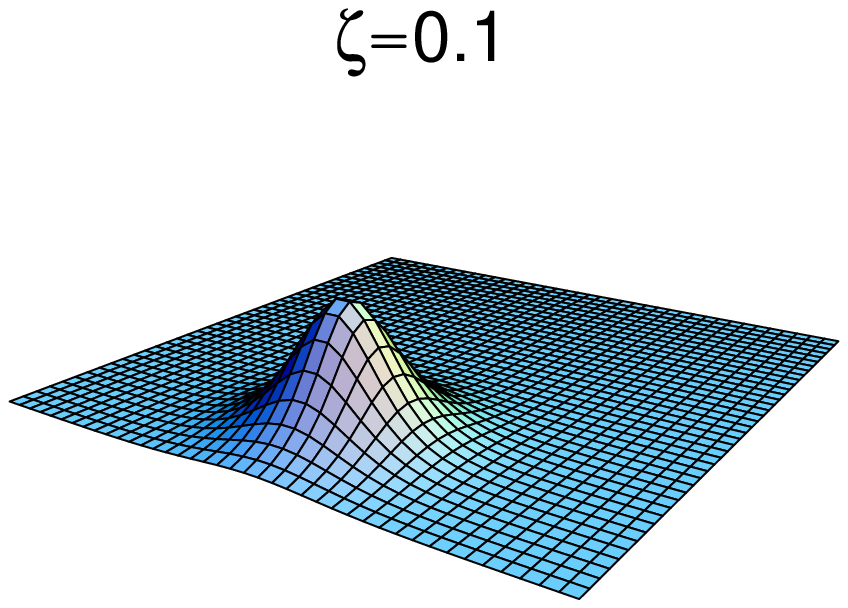,height=4cm,clip}
\hspace{-10mm}
\epsfig{file=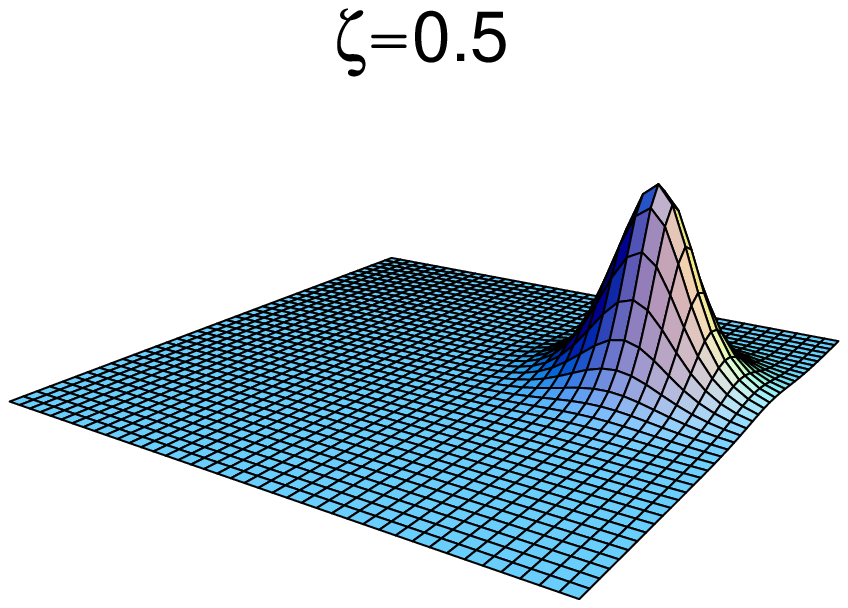,height=4cm,clip} \\
\vspace{5mm}
\epsfig{file=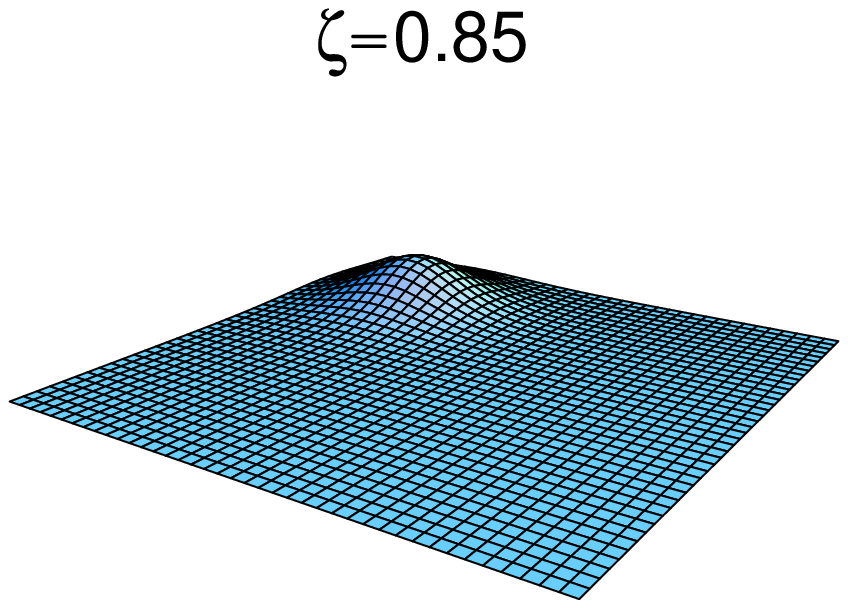,height=4cm,clip}
\end{center}
\caption{Scalar density of the KvB zero-modes for different
values of $\zeta$ in the $x,y$-plane. 
The parameters are the same as used in Fig.~\protect{\ref{f2example}},
i.e.~$(\mu_1,\mu_2,\mu_3) =
(-17,-2,19)/60$, $\vec{y}_1 = (-2,-2,0)$, $\vec{y}_2 = (0,2,0)$ and 
$\vec{y}_3 = (2,-1,0)$. 
\label{zmexample}}
\end{figure} 

Before we discuss the implications of this selection rule let us denote
a limiting case of the function $\hat{f}_{\zeta} (x)$. In the limit of 
well separated constituent monopoles, i.e.~large $|\vec{y}_i - 
\vec{y}_{i+1}|$ for all $i$ one finds
\begin{equation}
\hat{f}_{\zeta} (x) \; = \; 2\pi \, \frac{\sinh(2\pi(\zeta \! - \! \mu_m) r_m)
\sinh(2\pi(\mu_{m+1} \! - \! \zeta) r_m)}{r_m \, \sinh(2\pi\nu_m r_m)} \; .
\label{flimit}
\end{equation}
This limiting case together with the selection rule Eq.~(\ref{zmloc}) 
makes explicit the following two properties of the zero mode:

\begin{itemize}

\item The zero mode is located on the $m^{th}$ monopole when the
boundary condition parameter $\zeta$ is contained in $[\mu_m, \mu_{m+1}]$.
The position of the lump seen by the zero-mode can jump from 
$\vec{y}_m$ to $\vec{y}_{m+1}$ as the value of $\zeta$ crosses $\mu_{m+1}$.

\item The size of the zero-mode depends in addition to the
mass of the underlying monopole also on the parameter $\sigma$,
\begin{equation}
\sigma \; = \; \min \{\, \zeta-\mu_m \, , \, \mu_{m+1}-\zeta \, \} \; .
\label{sigma}
\end{equation}
The zero-mode is localized for large values of $\sigma$ and spread-out
for small $\sigma$.

\end{itemize}

In order to demonstrate the features of the zero-modes in Fig.\ 
\ref{zmexample} we show plots 
of the scalar density of the zero-modes for the example where we discussed  
the action density in the last subsection and displayed the
result in Fig.~\ref{f2example}. 
Our example has $(\mu_1,\mu_2,\mu_3,\mu_4) = (-17,-2,19,43)/60$.
This implies that for $\zeta \in [-2/60, 19/60]$ the zero mode
is located on $y_2$, for $\zeta \in [19/60, 43/60]$ it is located on
$y_3$ and for $\zeta \in [43/60, 58/60]$ it is located on $y_1$.
Note that we made use of the equivalence $\zeta \equiv \zeta -1$ in order 
to treat the interval $[\mu_1,\mu_2] = [-17/60,-2/60]$.

The plots show nicely that indeed the position of the lump seen by
the zero-mode can change its position.
In the following sections we now make use of this property and use 
the zero-modes of the lattice Dirac operator with different boundary 
conditions as a device for detecting different lumps that together
build up a topological excitation of charge $Q = \pm 1$.

\section{Finite Temperature - results above $T_c$}

\subsection{Overview over the configurations}

We remarked in Section 2 that the ensemble in the 
deconfined phase ($6\times20^3$, $\beta = 8.45$) consists of 
89 configurations, 25 with real 
Polyakov loop and 64 with complex Polyakov loop. 
For a subset of 10 configurations we did not only compute the zero-mode 
with periodic and anti-periodic temporal boundary conditions but for 
a general phase $\exp(i 2 \pi \zeta)$ at the temporal boundary. In particular 
we studied $\zeta =$ 0.0, 0.1, 0.2, 0.3, 1/3,
0.4, 0.5, 0.6, 2/3, 0.7, 0.8, 0.9. 
The cases $\zeta =$ 0.0 and $\zeta =$ 0.5 correspond to periodic 
respectively anti-periodic boundary conditions. For this subset 
Table~\ref{confs20b845} lists some basic properties. In particular we 
list the configuration number, the phase of the Polyakov loop, the inverse
participation ratio for the periodic and the anti-periodic zero-mode, 
the position of the maxima for periodic and anti-periodic b.c., and finally
the 4-distance between the two maxima.

\begin{table}[b!]
\begin{center}
\hspace*{-5mm}
\begin{tabular}{r|c|r|r|cccc|cccc|r}
conf. & $\varphi$ & $I_{P} $ & $I_{A} $ & 
\multicolumn{4}{c}{$(t,x,y,z)^{max}_{P}$} \vline &
\multicolumn{4}{c}{$(t,x,y,z)^{max}_{A}$} \vline & $d_4/a$ \\
\hline
6   & 0         &  2.19 &  43.16 & 5 & 10 &  9 &  8 & 3 &  4 & 17 &  9 & 10.24 \\
7   & $-2\pi/3$ & 22.81 &   4.71 & 3 &  3 &  3 & 11 & 1 &  4 &  2 &  9 &  3.16 \\
9   & $+2\pi/3$ & 23.70 &   5.93 & 5 &  8 & 18 & 17 & 5 &  8 & 18 & 17 &  0.00 \\
10  & $+2\pi/3$ & 40.27 &  10.04 & 1 & 11 & 13 & 14 & 1 & 11 & 13 & 14 &  0.00 \\
19  & 0         &  2.29 &  20.20 & 5 & 11 &  9 &  8 & 4 & 20 &  3 &  8 & 10.86 \\
38  & 0         &  2.38 &  31.00 & 5 & 12 &  9 &  8 & 5 & 12 &  9 &  8 &  0.00 \\
49  & $+2\pi/3$ & 34.03 &  15.67 & 6 & 11 & 19 &  2 & 6 & 11 & 19 &  2 &  0.00 \\
87  & 0         &  1.90 &  47.76 & 6 & 18 & 13 &  9 & 6 & 18 & 13 &  9 &  0.00 \\
101 & 0         &  2.87 &  41.69 & 5 & 12 & 18 & 10 & 1 & 19 &  3 &  1 & 12.60 \\
383 & 0         &  2.42 &  42.86 & 6 & 19 & 13 & 11 & 2 &  3 & 11 &  4 &  8.54 
\end{tabular}
\end{center}
\caption{Parameters of the subensemble in the deconfined phase 
($6\times20^3$, $\beta = 8.45$)
which we studied with finely spaced values for the boundary parameter $\zeta$. 
For these 10 configurations 
we quote the configuration number,
the phase of the Polyakov loop, the inverse participation ratios 
$I_{P}$ and $I_{A}$ for the zero-modes with periodic, 
respectively anti-periodic b.c., the position of the maxima for
the corresponding zero-modes and the 4-distance $d_4$ between these two 
maxima in units of the lattice spacing.
\label{confs20b845}}
\end{table}

A few basic properties of the configurations above $T_c$ can be immediately
read off from Table~\ref{confs20b845}. For all configurations with 
real Polyakov loop ($\varphi = 0$) the inverse participation ratio 
of the zero-mode increases when switching from periodic to anti-periodic 
boundary conditions, i.e.~the mode becomes more localized. For 
the configurations with complex Polyakov loop ($\varphi = \pm 2\pi/3$)
the situation is reversed and the state becomes more delocalized when switching
from periodic to anti-periodic b.c. Furthermore only configurations with 
real Polyakov loop allow for a large distance between the lumps seen by the
periodic and anti-periodic zero mode. In the following subsections
we will elaborate further on these observations.

\subsection{Example for the generic behavior}

In this subsection we show series of plots of the scalar density for
a configuration where we computed with the zero-modes with finely 
spaced values of $\zeta$. Before we do so let us discuss what can be 
expected from the analytic KvB result.

In the deconfined phase the Polyakov loop can have the
three phases $\varphi = 0, \pm 2\pi/3$. Let us first discuss
the case of real Polyakov loop. For this case one finds for the phase 
factors $\mu_i$ entering the KvB solutions,
\begin{equation}
(\mu_1,\mu_2,\mu_3,\mu_4) \; \; = \; \; (0,0,0,1) \; .
\label{muforzero}
\end{equation}
Applying the rule that the zero mode with boundary factor $\zeta$
is located at position $\vec{y}_m$ when $\zeta \in [\mu_m,\mu_{m+1}]$
we find: For all $\zeta \neq 0$ the zero mode is located 
on $\vec{y}_3$. The only exception is $\zeta = 0$ where the zero mode
is located on $\vec{y}_1$ and $\vec{y}_2$. Note that $\zeta = 0$
is equivalent to  $\zeta = 1$ 
such that also for $\zeta = 1$ this second possibility 
holds. The mode at $\zeta = 0,1$ should be very much delocalized, while
the other mode seen for $\zeta \neq 0$ is expected to show a changing 
degree of localization. In particular one expects a rather delocalized 
mode near the endpoints of the interval $\zeta = 0,1$ while
it is more localized in the middle of the interval, reaching a maximum
of the localization at $\zeta = 0.5$. 

The plots we will show below are for our configuration 383 from
the $6\times20^3, \beta = 8.45$ ensemble. In Table \ref{383max} we show
the inverse participation ratio and the position of the maximum
of the scalar density for all values of $\zeta$. From the table it is obvious
that the mode is located at the same position for all $\zeta \neq 0,1$
and at a different position for $\zeta = 0,1$. The distance between the 
two lumps is $d_4/a = 8.54$ in lattice units or in fermi 
$d_4 = 0.80$ fm. The inverse 
participation ratio is largest near $\zeta = 0.5$, i.e.~the mode is most 
localized for anti-periodic boundary conditions. At $\zeta = 0,1$
the inverse participation ratio is small and the mode is very much spread 
out. 

\begin{table}[t]
\begin{center}
\hspace*{-5mm}
\begin{tabular}{l|c|cccc|l|c|cccc}
$\;\;\zeta$ & $I$ & \multicolumn{4}{c}{$(t,x,y,z)^{max}$} \vline & $\;\;\zeta$ & $I$ &
\multicolumn{4}{c}{$(t,x,y,z)^{max}$} \\
\hline
0.0  &  2.42 & 6 &  9 & 13 & 11 &  0.6  & 45.32 & 2 & 13 & 11 &  4  \\
0.1  &  3.53 & 2 & 13 & 11 &  4 &  0.7  & 35.20 & 2 & 13 & 11 &  4  \\
0.2  &  9.15 & 2 & 13 & 11 &  4 &  0.8  & 17.00 & 2 & 13 & 11 &  4  \\
0.3  & 19.05 & 2 & 13 & 11 &  4 &  0.9  &  3.40 & 2 & 13 & 11 &  4  \\
0.4  & 31.80 & 2 & 13 & 11 &  4 &  1.0  &  2.42 & 6 &  9 & 13 & 11  \\
0.5  & 42.86 & 2 & 13 & 11 &  4 &       &       &   &    &    &  
\end{tabular}
\end{center}
\caption{Inverse participation ratio and position of the maximum of
the scalar density as a function of the boundary condition parameter 
$\zeta$. The data are for configuration 383 of the deconfined ensemble
($6\times20^3, \beta = 8.45$). Plots of the corresponding scalar density 
are shown in Fig.\ \protect{\ref{dens20b845_383a}}.
\label{383max}}
\end{table}

\begin{figure}[t!]
\begin{center}
\epsfig{file=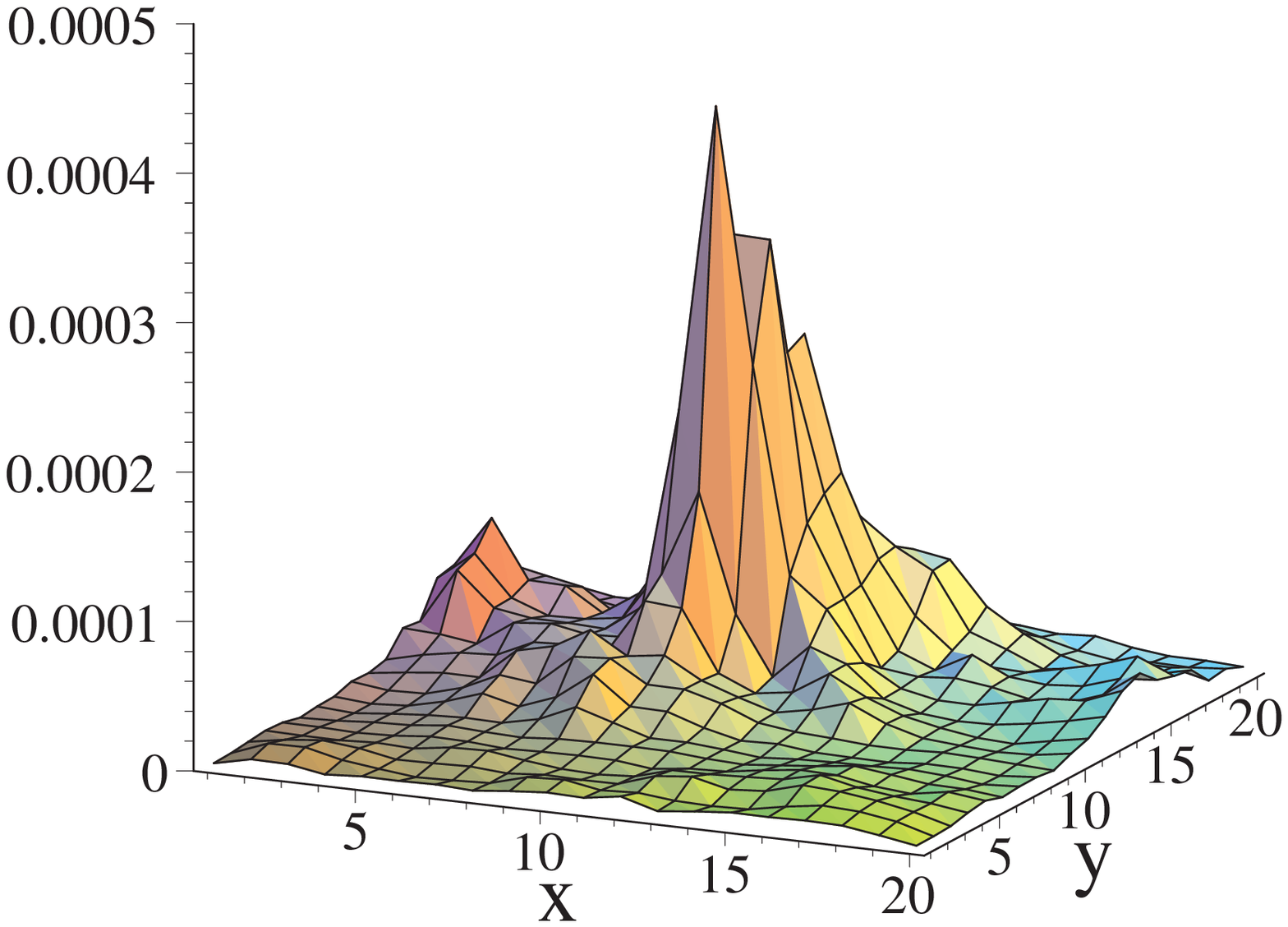,height=3.5cm,clip}
\epsfig{file=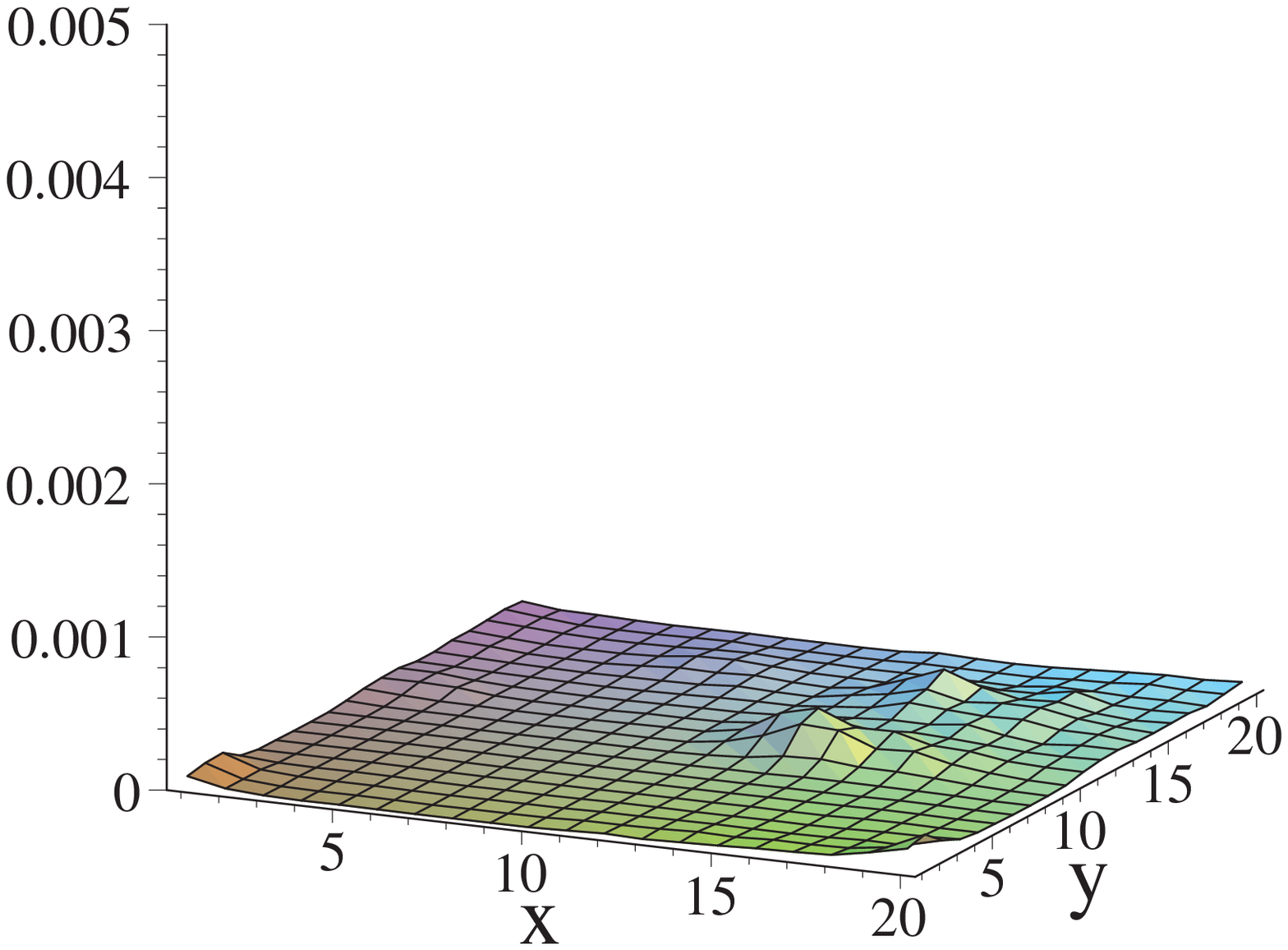,height=3.5cm,clip}
\\
\epsfig{file=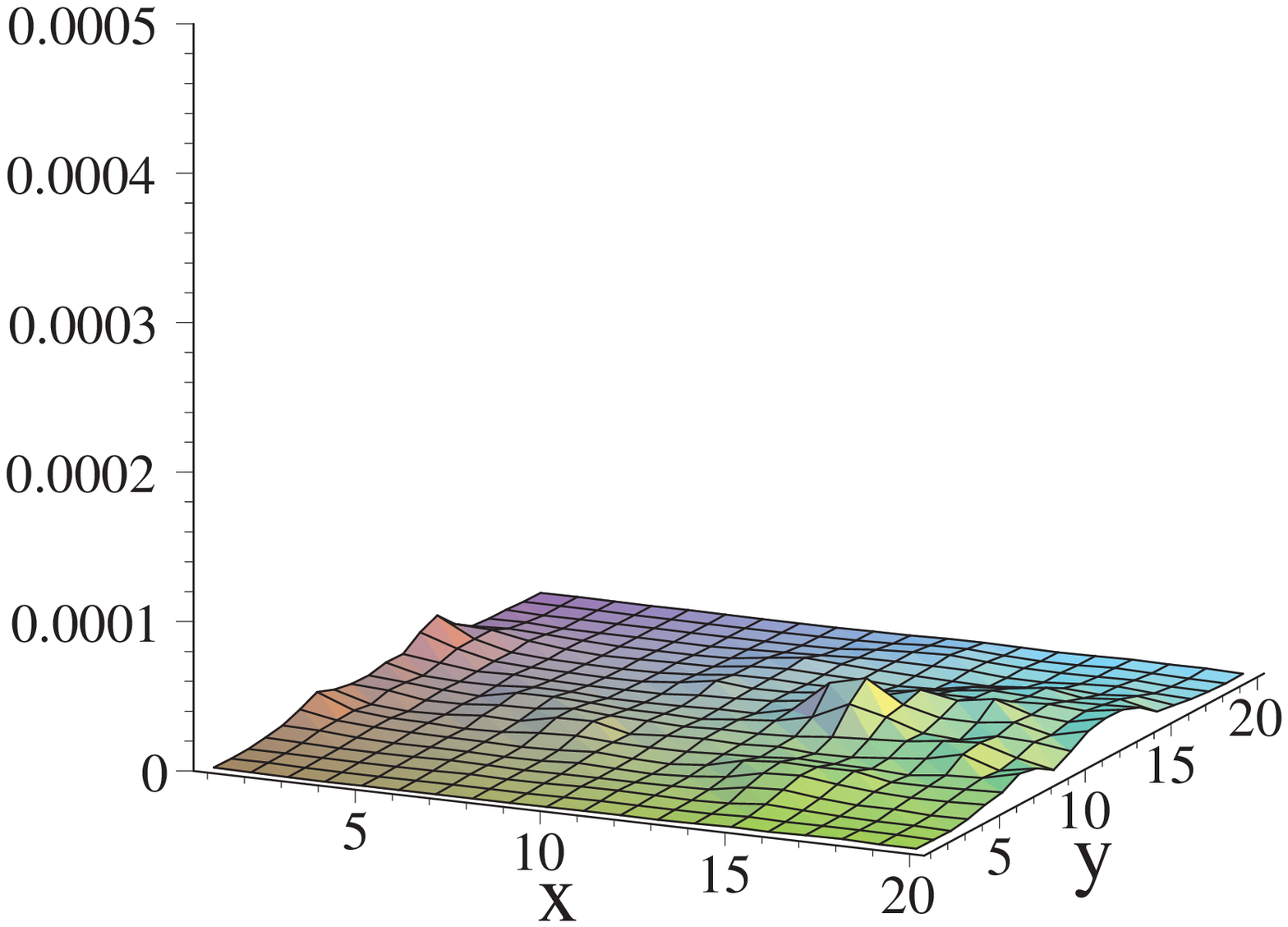,height=3.5cm,clip}
\epsfig{file=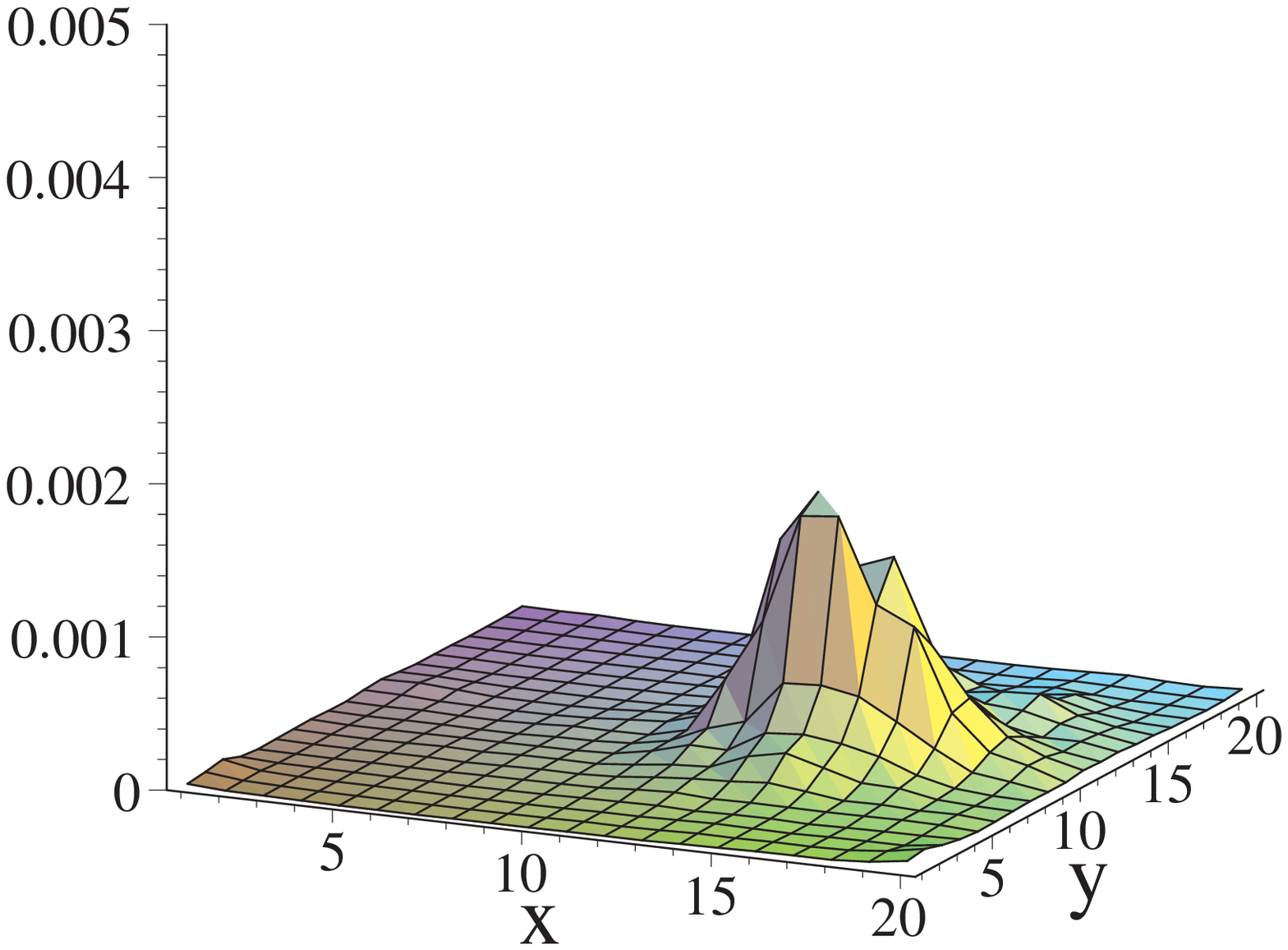,height=3.5cm,clip}
\\
\epsfig{file=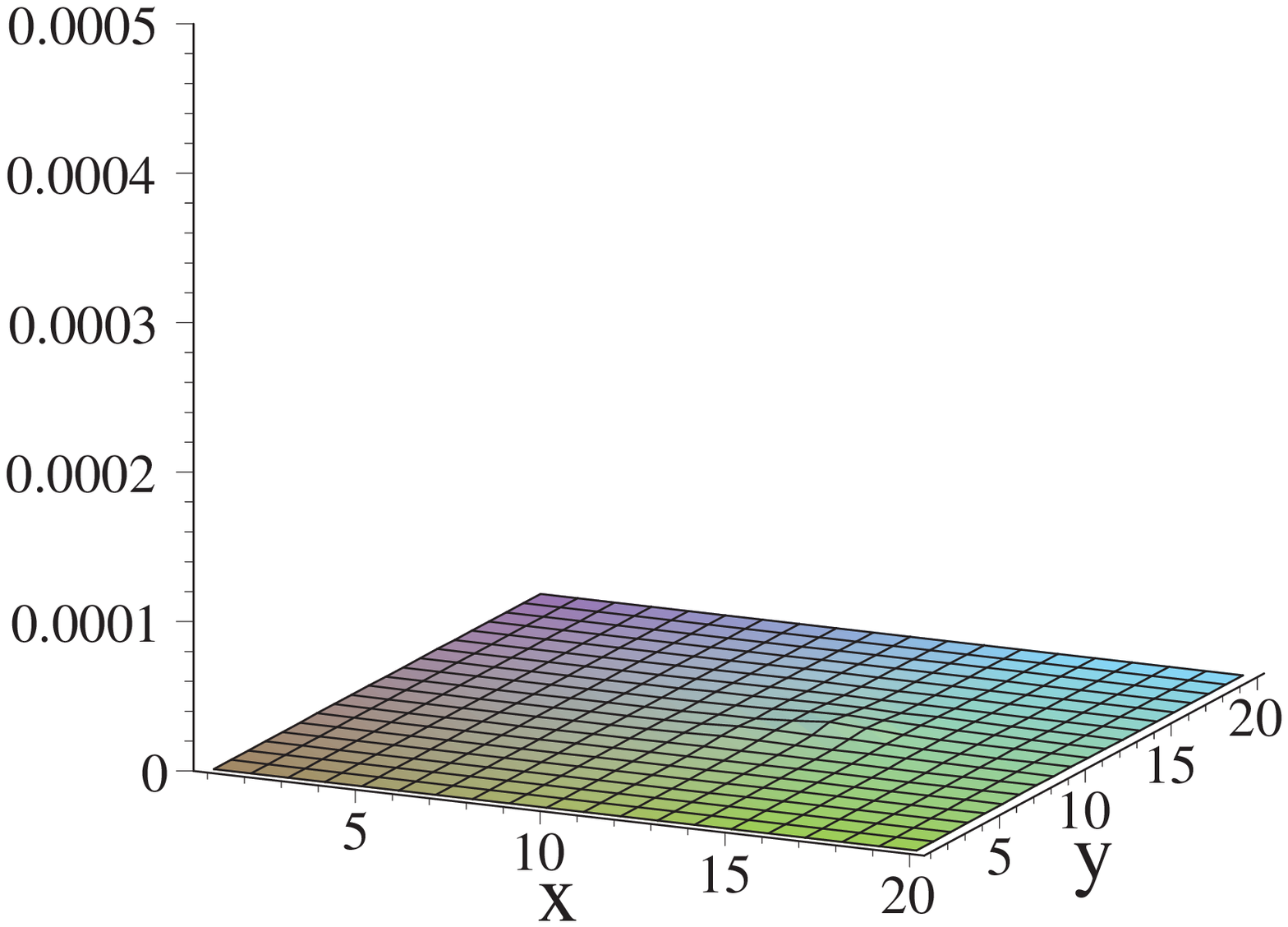,height=3.5cm,clip}
\epsfig{file=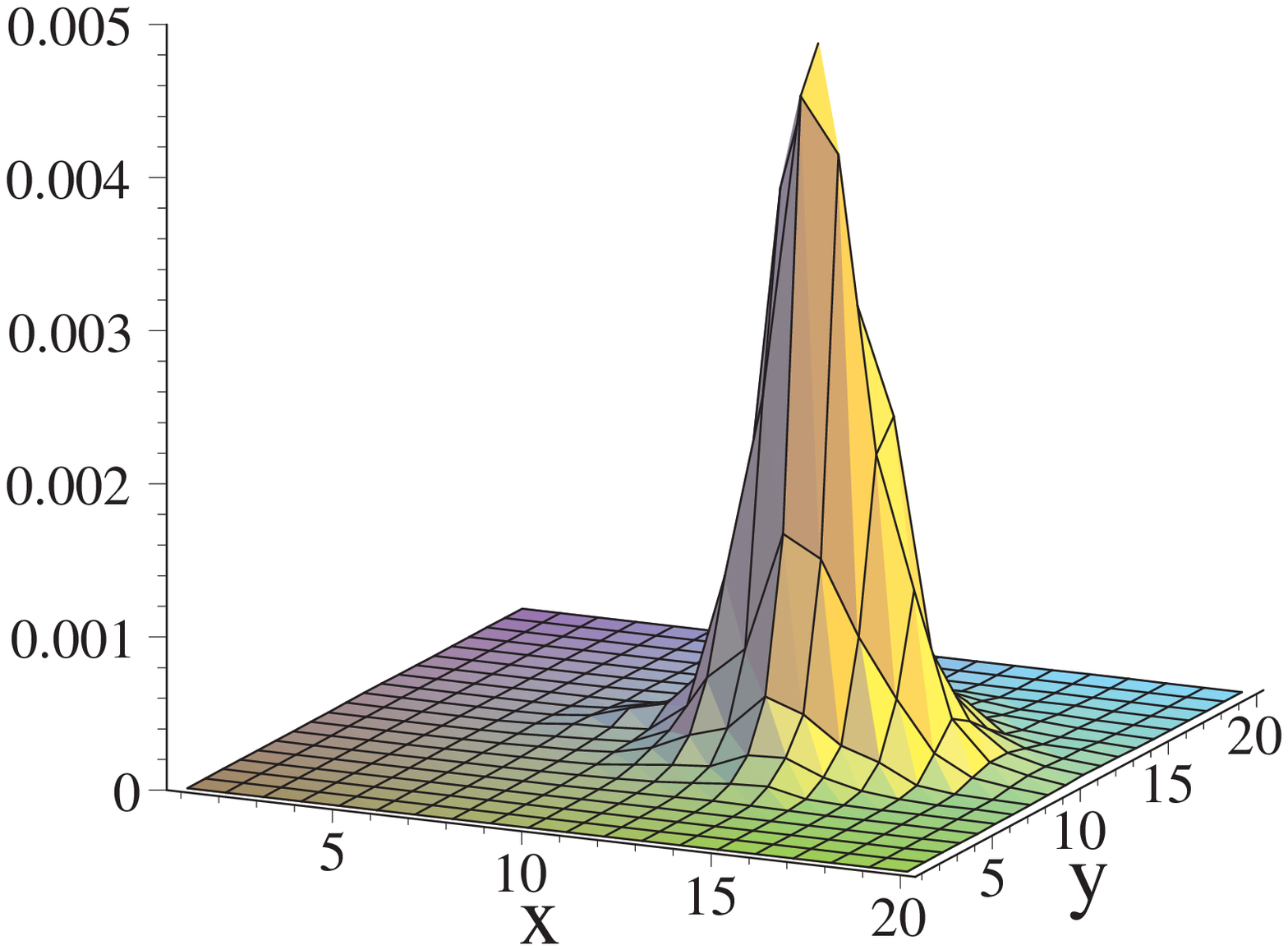,height=3.5cm,clip}
\end{center}
\vspace{5mm}
\caption{Slices of the scalar density for $6\times20^3, 
\beta=8.45$ and real Polyakov loop (configuration 383). 
We show $x,y$-slices at 
$t = 6$, $z = 11$ (left column)  and at 
$t = 2$, $z = 4$ (right column). The values for $\zeta$ are
$\zeta = 0, 0.2, 0.5$ (from top to bottom). 
Note the different scale for the l.h.s.~and r.h.s.~plots.
\label{dens20b845_383a}}
\end{figure}

Let us now look at the plots of the scalar density for configuration 
383. In Fig.~\ref{dens20b845_383a} we show 
the scalar density on $x,y$-slices of the lattice. In particular we
show the slice at $t,z = 6,11$ in the left column of plots and 
the slice at $t,z = 2,4$ in the right column of plots. Thus in
the left column we show the slice where the scalar density at 
$\zeta = 0,1$ has its peak while in the right column the slice
for the maximum at all other values of $\zeta$ is shown. The reason 
for showing two different slices is that typically for a thermalized 
configuration the positions of the two maxima do not fall 
in a common coordinate plane. In the plots we use
values of $\zeta = 0,0.2$ and 0.5. Note that the scale for the 
right column differs by a factor of 10 from the scale in the 
left column. 

For $\zeta = 0$ there is a clearly visible lump in the 6,11-slice
(left column, top plot of Fig.~\ref{dens20b845_383a}), 
while the 2,4-slice is essentially flat
(right column, top plot). As we increase $\zeta$ to $\zeta = 0.2$ the 
peak in the 6,11-slice has vanished, but a small lump in the 
2,4-slice has emerged (second row in Fig.~\ref{dens20b845_383a}).
This latter lump starts to grow as we increase $\zeta$ and
reaches its maximum in the last row of Fig.~\ref{dens20b845_383a},
at $\zeta = 0.5$. As we increase $\zeta$ further the lump starts to shrink 
again, and essentially runs through the series of plots in reverse order.
The series of plots shown in 
Fig.~\ref{dens20b845_383a}
nicely demonstrates that the zero-mode for configuration 383 behaves 
exactly like the zero-mode for a KvB solution in a deconfined 
configuration with real Polyakov loop.

Let us now look at the 
other two possibilities for the phase of the Polyakov loop. 
Phases of $\varphi = 2\pi/3$ and $\varphi = -2\pi/3$ correspond
to the following two sets of $\mu_i$,
\begin{equation}
(\mu_1,\mu_2,\mu_3,\mu_4) \; \; = \; \; (-2/3,1/3,1/3,1/3) \; ,
\label{muforplus}
\end{equation}
for $\varphi = 2\pi/3$ and
\begin{equation}
(\mu_1,\mu_2,\mu_3,\mu_4) \; \; = \; \; (-1/3,-1/3,2/3,2/3) \; ,
\label{muforminus}
\end{equation}
for $\varphi = -2\pi/3$.
Thus for $\varphi = 2\pi/3$ the zero-mode has to be located 
at the same position for all $\zeta \neq 1/3$ and only 
for $\zeta \in [1/3,1/3]$ (in other words for $\zeta = 1/3$) the lump
can sit at a different position. For $\varphi = 2\pi/3$ the zero-mode 
has to be located 
at the same position for all $\zeta \neq 2/3$, but can have a different 
location for $\zeta = 2/3$. 
This discussion shows that the situation for $\varphi = \pm 2\pi/3$ 
is equivalent to the $\varphi = 0$ case and only the interval $[0,1]$
is shifted, such that the position of a possible jump of the 
location is shifted to $\zeta = 1/3$, respectively $\zeta = 2/3$.

We find that all our configurations in the deconfined ensemble
obey the general pattern we have discussed. Certainly it is also
clear that our thermalized configurations are not classical solutions
and quantum effects play a role. For example we find for the 
configuration shown in Fig.~\ref{dens20b845_383a}
that the peak in the $6,11$-slice is already visible at $\zeta = 0.9$. 
For an unperturbed KvB solution it
should be visible only at exactly $\zeta = 1$ (equivalent to $\zeta = 0$).
The quantum effects seem to slightly twist the Polyakov loop 
such that it is not exactly an element of the center. This makes 
the critical values of $\zeta = 0,1/3,2/3$ where the 
lump can change its position less strict than for the continuum
solution.

\subsection{Results from the whole sample}
In the last subsection we have discussed an example for the generic
behavior of the KvB zero-modes in the deconfined phase and remarked 
that all configurations we looked at show the same general pattern.
Here we now present results for the whole set of configurations 
listed in Table~\ref{confs20b845} where we used finely spaced 
$\zeta$ and also for the even larger set of 89 configurations where
we only compared periodic and anti-periodic boundary conditions. 

\begin{figure}[t!]
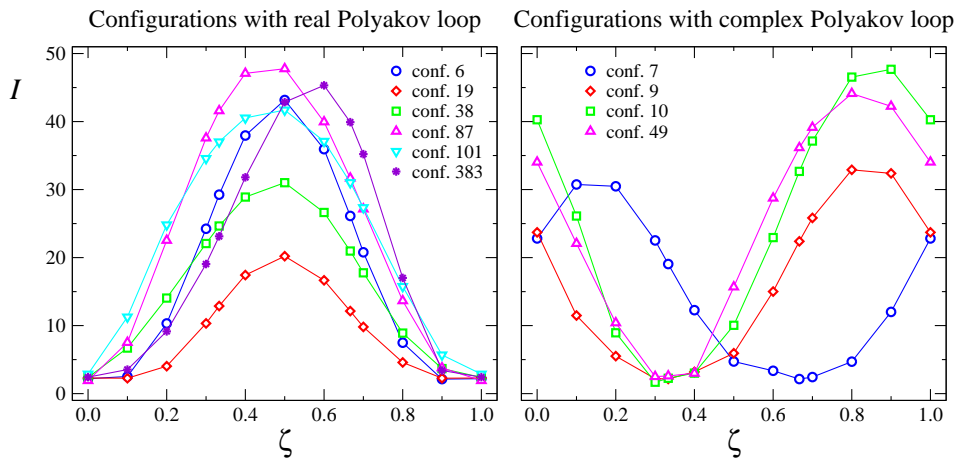

\begin{center}
\hspace*{-3mm}
\epsfig{file=ipr_vs_phase_20_b845_real.eps,height=6cm,clip}
\hspace{-1mm}
\epsfig{file=ipr_vs_phase_20_b845_complex.eps,height=6cm,clip}
\end{center}
\vspace{-3mm}
\caption{Inverse participation ratio $I$ as a function of $\zeta$. 
We show the results for the deconfined phase ($6\times20^3, \beta = 8.45$), 
displaying configurations 
with real Polyakov loop in the l.h.s.~plot and configurations
with complex Polyakov loop in the r.h.s.~plot.
\label{ipr_vs_phase_20_b845}}
\end{figure}

Let us begin with looking at the behavior of the inverse participation
ratio, i.e.~the localization property. The discussion in the
last section and the example shown there suggest that for all 
configurations in the real sector the inverse participation ratio has a 
maximum near $\zeta = 0.5$ (the zero-mode is most localized there) 
and a minimum near $\zeta = 0,1$. For configurations with complex 
Polyakov loop the whole curve is shifted such that the minimum is
at $\zeta = 1/3$ for $\varphi = 2\pi/3$ and at $\zeta = 2/3$
for $\varphi = -2\pi/3$. Note that this behavior is independent from a 
possible change of the position of the peak. 

In Fig.~\ref{ipr_vs_phase_20_b845} we show for all configurations
in the deconfined phase where we did runs with small steps for $\zeta$
how the localization changes as a function of $\zeta$. The l.h.s.~plot gives 
the results for configurations with real Polyakov loop, while the 
r.h.s.~plot is for complex Polyakov loop. 

The plots show that indeed for all these configurations 
the predicted pattern holds.
The configurations with real Polyakov loop (l.h.s. plot)
have a maximum of $I$ near 
$\zeta = 0.5$ and minima near 0 and 1. For configurations with complex 
Polyakov loop 
(r.h.s.\ plot) we find the curve shifted such that the minimum is
at $\zeta = 1/3$ for phase $\varphi = 2\pi/3$ (configurations 9, 10, 49 in the
r.h.s.\ plot) and at $\zeta = 2/3$ for phase $\varphi = -2\pi/3$ 
(configuration 7). 
 
From the above discussions of the properties of KvB zero-modes 
and the examples we gave one can derive another prediction. This
prediction links the distance between the positions of the lumps
in the zero-mode with periodic and anti-periodic boundary
conditions to the phase of the Polyakov loop: For real Polyakov loop
the periodic mode with $\zeta = 0$ can sit in either the interval 
$[0,0]$ or in the interval $[0,1]$, while the periodic zero-mode
with $\zeta = 0.5$ is only contained in $[0,1]$. Thus for real Polyakov 
loop the periodic and the anti-periodic zero-mode can be located at 
different positions. The situation is different for complex Polyakov loop.
E.g.~for $\varphi = -2\pi/3$ both $\zeta = 0$ and $\zeta = 0.5$ 
are contained in the interval $[-1/3,2/3]$ and the periodic as well as
the anti-periodic zero-mode are expected to be located at the same position.

\begin{figure}[t]
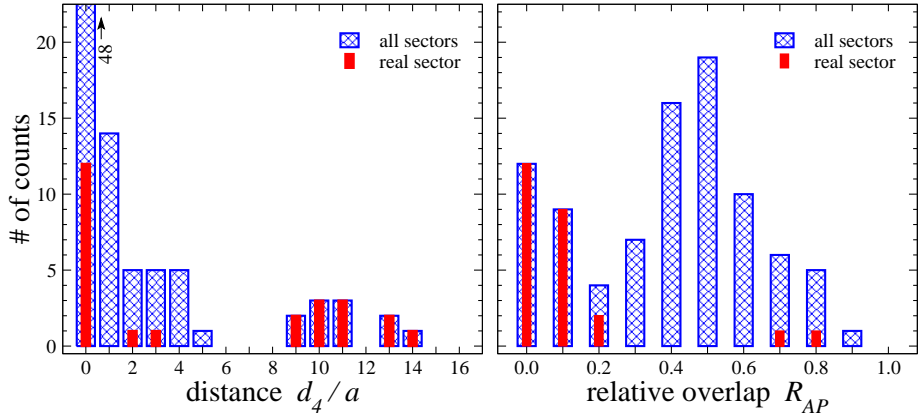

\begin{center}
\vspace{5mm}
\hspace*{-5mm}
\epsfig{file=disthisto6x20b845.eps,height=5.5cm,clip}
\epsfig{file=overlaphisto6x20b845.eps,height=5.5cm,clip}
\end{center}
\vspace{-3mm}
\caption{Histograms for the distance $R_{AP}$
between the peaks in the scalar density 
when comparing periodic to anti-periodic b.c.\ (l.h.s.\ plot) 
and for the corresponding 
relative overlap $R_{AP}$ (r.h.s.\ plot). The data are for the 
deconfined ensemble on $6\times20^3$ lattices at $\beta = 8.45$.
\label{disthisto20b845}}
\end{figure}

\begin{figure}[t]
\begin{center}
\vspace{5mm}
\hspace*{-5mm}
\epsfig{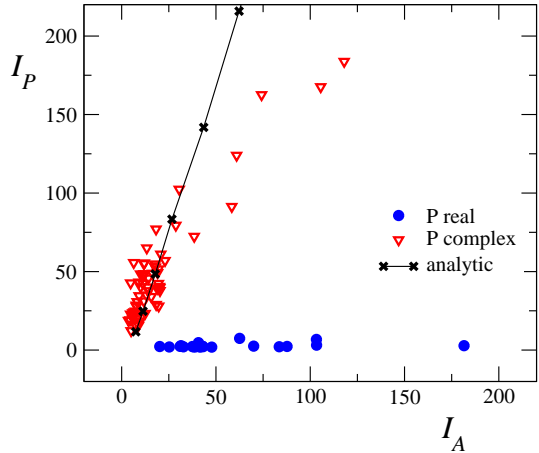}
\end{center}
\vspace{-3mm}
\caption{Scatter plot for the inverse participation ratio of the
zero mode with periodic and anti-periodic boundary conditions. 
The data are for the deconfined ensemble on $6\times20^3$ 
lattices at $\beta = 8.45$. We use filled circles for configurations with 
real Polyakov loop and open triangles for the complex sector. 
We also include numbers (crosses connected with straight lines to
guide the eye) that were generated from the analytic solution.   
\label{iprscatter}}
\end{figure}

In Fig.~\ref{disthisto20b845} we test this property and in the l.h.s.\ plot
show histograms of the distance between the
peaks in the scalar density when comparing periodic to anti-periodic 
temporal boundary conditions. 
From the histogram it is obvious that indeed all configurations with large 
distances between the peaks, i.e.~distances larger than 5 lattice spacings
(0.47 fm) have real 
Polyakov loop.
About a third of the configurations with complex Polyakov loop
also show small but non-zero distances (up to distance
5 for a single configuration)
between the periodic and the anti-periodic lump. We believe that this
is due to quantum fluctuations superimposed on the lumps.
Note that also for real Polyakov loop the KvB zero modes
can sit at the same position a case which is realized by about half of the
zero modes in the real sector. The plot on the r.h.s.\ of
Fig.~\ref{disthisto20b845} shows histograms for the relative overlap $R_{AP}$
of the periodic and the anti-periodic zero mode.
These results corroborate the findings of the histograms for the distance
since the configurations which have no or only small overlap are all in the
real sector. Note that a relative overlap smaller than 1 can also
come from two modes which sit at the same position but have different
localization. This explains why also configurations in the complex
sector have relative overlap smaller than 1 (compare also
Fig.~\ref{iprscatter} and the corresponding discussion below).

Finally in Fig.~\ref{iprscatter} we show a scatter plot of the 
values of the inverse participation ratios $I_P, I_A$ 
for periodic and 
anti-periodic boundary conditions. For the real sector 
the mode is delocalized and $I_P$ is small. Thus one expects
that in the $I_P$ versus $I_A$ scatter plot the data fall on a horizontal 
line. Our numerical results clearly show this behavior. For the complex 
sector the periodic mode has a scale parameter $\sigma \sim 1/3$ while
the anti-periodic zero mode has $\sigma \sim 1/6$. Thus one expects 
$I_P > I_A$ for the complex sector and again our data nicely confirm this
prediction. The plot also contains numbers of $I_P$ and $I_A$ that were 
generated from the analytic solution for complex Polyakov loop. 
We varied the positions $\vec{y}_i$ of the monopoles thus obtaining
different values for the inverse participation ratio. These data are 
shown as crosses in the plot and we connect them to guide the eye. 
Our numerical results in the complex sector follow this analytic curve  
quite well. Only for very strongly localized modes ($I_P > 100$) 
we see a slight
deviation. We remark that on the lattice such strongly localized states
are resolved by only a few lattice points such that cut-off effects 
become important.

\subsection{Results for the spectral gap}

The chiral condensate is related to the spectral density $\rho_{spec}(\lambda)$
of the Dirac operator near the origin through the Banks-Casher formula
\cite{bankscasher},
\begin{equation}
\langle\overline{\psi} \psi \rangle \; \; = \; \; - \, 
\lim_{V\rightarrow \infty} \, \frac{\pi}{V} \, \rho_{spec}(0) \; .
\end{equation}
For the confining phase where chiral symmetry is broken the spectral 
density extends all the way to the origin. In the deconfined phase where
chiral symmetry is restored a gap opens up in the spectral density near 
the origin. $\rho_{spec}(0)$ vanishes and so does the condensate. 
The opening up of the spectral gap is believed to come from an arrangement 
of instantons and anti-instantons to tightly bound molecules \cite{instantons}. 

An interesting question is whether the chiral condensate vanishes at the
same temperature for all sectors of the Polyakov loop. Results \cite{ChCh95}
for staggered fermions with anti-periodic 
temporal b.c.~seemed to indicate that the chiral
condensate vanishes at different temperatures for real and complex
Polyakov loop, and attempts to understand this phenomenon can be found
in \cite{subsequent}. A more recent study \cite{crittemp}
analyzing directly the spectral gap (also with anti-periodic temporal b.c.)
came to a different conclusion: The
spectral gap appears at the same critical temperature for all sectors 
of the Polyakov loop and this temperature coincides with the
deconfinement transition. 
Here we further corroborate the scenario of a single transition
by discussing the
spectral gap for different boundary conditions for the Dirac 
operator.
\begin{figure}[t]
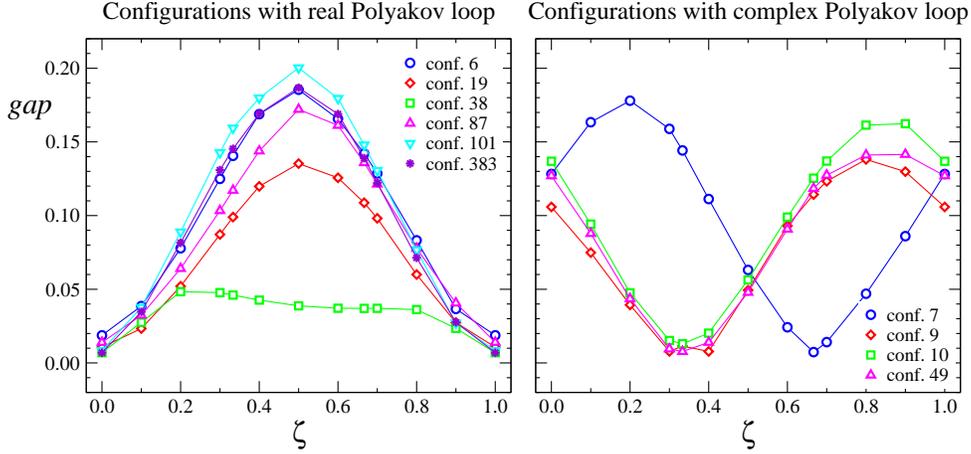

\begin{center}
\hspace*{-3mm}
\epsfig{file=gap_vs_phase_b845_real.eps,height=6cm,clip}
\hspace{-1mm}
\epsfig{file=gap_vs_phase_b845_complex.eps,height=6cm,clip}
\end{center}
\vspace{-3mm}
\caption{Size of the spectral gap as a function of $\zeta$. 
We show the results for the deconfined phase, displaying configurations 
with real Polyakov loop in the l.h.s.~plot and configurations
with complex Polyakov loop in the r.h.s.~plot.
\label{gap_vs_zeta}}
\end{figure}

In Fig.~\ref{gap_vs_zeta} we show the spectral gap as a function
of the boundary condition parameter $\zeta$ for the 10 
configurations in the deconfined phase where spectra with
finely spaced $\zeta$ were computed. The l.h.s.\ plot shows
the 6 configurations with real Polyakov loop, while the r.h.s\ plot 
gives data for the configurations with complex Polyakov loop.
The spectral gap is defined as the size of the imaginary part of the 
smallest complex eigenvalue. 

The plots show that as for the inverse participation ratio the results
for the gap from configurations with $\varphi = \pm 2\pi/3$ for the Polyakov loop 
can be obtained from the result for $\varphi = 0$ by shifting the 
$\zeta$ axis by amounts of $1/3$ and $2/3$ respectively. The results in
the three sectors are completely equivalent (with only the data for 
configuration 38 showing a somewhat irregular behavior) 
and can be transformed into
each other through different boundary conditions for the fermions. 
This symmetry shows that there is no reason to expect a different
critical temperature for chiral symmetry restoration in different 
sectors of the Polyakov loop. 

It is interesting to note that in the real sector for $\zeta = 0$ and in 
the complex sectors for $\zeta = 1/3$, respectively $\zeta = 2/3$
the spectral gap becomes very small. This could indicate that for
these boundary conditions chiral symmetry remains broken also
in the deconfined phase. It would be interesting to analyze the 
chiral condensate in the deconfined phase using these boundary conditions
for the fermions.

\section{Finite Temperature - results below $T_c$}

\subsection{Overview over the configurations}

In the last section we have studied the behavior of the configurations 
in the deconfined phase. We have found that qualitatively our results 
match the characteristics of the KvB solutions very well. 
In this section we now turn to the ensemble below $T_c$. Below $T_c$ 
the Polyakov loop vanishes which leaves only the unique possibility,
\begin{equation}
(\mu_1,\mu_2,\mu_3,\mu_4) \; \; =  \; \; (-1/3, 0, 1/3, 2/3) \; ,
\end{equation}
for the phase
factors $\mu_i$. One expects that the zero-modes can change 
their position when $\zeta$ crosses the values of $1/3$ and $2/3$. Thus below
$T_c$ the zero-mode can change its position more often and for 
these configurations we now use a finer spacing of $\zeta$. 

\begin{table}[t!]
\begin{center}
\hspace*{-5mm}
\begin{tabular}{r|r|r|cccc|cccc|r|c}
conf.& $I_{P} $ & $I_{A}$ 
& 
\multicolumn{4}{c}{$(t,x,y,z)^{max}_{P}$} \vline & 
\multicolumn{4}{c}{$(t,x,y,z)^{max}_{A}$} \vline & $d_4/a$ & \# lumps\\
\hline
13  &  4.67 &  13.41 & 5 & 11 & 16 & 11 & 2 &  1 &  9 &  6 & 13.52 & 4 \\
18  &  7.32 &  10.73 & 2 &  6 & 20 &  3 & 6 & 16 & 15 & 12 & 14.49 & 3 \\
92  & 38.52 &   5.73 & 1 & 11 & 13 & 13 & 4 &  8 & 16 & 11 &  5.56 & 2 \\
109 &  5.56 &  15.93 & 1 & 14 & 18 & 18 & 2 & 18 &  5 &  3 &  9.53 & 4 \\
123 &  4.59 &  32.21 & 5 & 18 & 19 & 11 & 4 & 13 & 18 &  1 & 11.26 & 2 \\
125 &  4.13 &  14.72 & 5 &  1 & 16 &  9 & 5 & 14 & 16 & 18 & 11.40 & 3 \\
153 & 22.98 &  10.32 & 5 & 17 & 18 & 19 & 6 & 18 & 17 &  2 &  3.46 & 2 \\
210 &  2.48 &  28.96 & 1 &  1 &  8 &  4 & 2 &  2 &  1 & 13 & 11.48 & 3 \\
215 &  7.23 &   3.63 & 1 & 15 & 20 & 14 & 5 & 11 & 17 &  6 &  9.64 & 3 \\
266 &  8.15 &  40.01 & 1 &  9 & 19 & 20 & 2 &  9 &  6 &  6 &  9.27 & 3 
\end{tabular}
\end{center}
\caption{Parameters of the subensemble below $T_c$
($6\times20^3, \beta = 8.20$) which we studied
with finely spaced values for the boundary parameter $\zeta$. 
For these 10 configurations 
we list the configuration number,
the inverse participation ratios 
$I_{P}$ and $I_{A}$ for the zero-modes with periodic, 
respectively anti-periodic b.c., the position of the maxima for
the corresponding zero-modes, the 4-distance $d_4$ between these two 
maxima and the number of different lumps visited by the mode 
in a complete cycle through $\zeta$.
\label{confs20b820}}
\end{table}

From the complete ensemble below $T_c$ (compare Table \ref{confdata})
we computed for 10 configurations eigenvectors at all values of $\zeta$
between $\zeta = 0$ and $\zeta = 1$ in steps of 0.05. In 
Table \ref{confs20b820} we list some basic properties
of these configurations. In particular we list the configuration number,
the inverse participation ratios 
$I_{P}$ and $I_{A}$ for the zero-modes with periodic, 
respectively anti-periodic b.c., the position of the maxima for
the corresponding zero-modes and the 4-distance $d_4$ between these two 
maxima in lattice units. 
Finally we list the number of different lumps visited by the mode 
in a complete cycle through $\zeta$. We count a lump as independent lump
when its maximum remains at the same position for at least 3 subsequent
values of $\zeta$, i.e.~dominates the zero mode at least for 
a $\zeta$-interval of length 0.1. Note that these lumps are visited by the 
zero-mode one after the other and typically one of them dominates the 
zero-mode.

A first interesting finding can already be read off from the 
numbers in the table. 
As discussed, below $T_c$ the zero mode has three possible values
of $\zeta$ where it can change position, $\zeta = 0, 1/3, 2/3$. Thus
the zero-mode can visit up to 3 different positions. Also a number less than 
3 is possible since the coordinate vectors $\vec{y}_i$ of the monopoles
can coincide. Most of the eigenmodes do indeed see 2 or 3 lumps but for 
two of the configurations (13 and 109) we found 4 lumps. This observation
indicates that for our ensemble below $T_c$ the zero-modes do not
follow the predictions for KvB zero-modes as closely as 
their counterparts above $T_c$. Reasons for that may be quantum 
effects which drive the trace of the Polyakov loop away from zero.
We also remark that 
the lattice below $T_c$ is coarser than the lattice above $T_c$.

\subsection{Example for the generic behavior}

As for the ensemble in the deconfined phase we begin the presentation
of our results with showing slices of the scalar density
for a generic example. As discussed
in the last subsection, below $T_c$ the zero-mode is expected to 
visit up to three different lumps in a complete cycle through the
boundary condition. Configuration 125 is an example for such a behavior. 
In Table \ref{125table} we list some basic properties of this configuration
for all values of $\zeta$ we studied. 

\begin{table}[b!]
\begin{center}
\hspace*{-5mm}
\begin{tabular}{l|c|cccc|l|c|cccc}
$\;\;\zeta$ & $I$ & 
\multicolumn{4}{c}{$(t,x,y,z)^{max}$} \vline &
$\;\;\zeta$ & $I$ & 
\multicolumn{4}{c}{$(t,x,y,z)^{max}$} \\
\hline
 0.0   &  4.13 & 5 & 13 &  6 &  9 & 0.55 & 16.61 & 5 &  6 & 6 & 18  \\
 0.05  &  5.70 & 5 & 13 &  6 &  9 & 0.6  & 18.36 & 5 &  6 & 6 & 18  \\  
 0.1   &  6.03 & 5 &  6 &  6 & 18 & 0.65 & 18.07 & 5 &  6 & 6 & 18  \\
 0.15  &  6.26 & 5 &  6 &  6 & 18 & 0.7  & 16.17 & 5 &  6 & 6 & 18  \\ 
 0.2   &  5.50 & 2 &  6 & 13 & 19 & 0.75 & 13.10 & 5 &  6 & 6 & 18  \\
 0.25  &  5.60 & 2 &  6 & 13 & 19 & 0.8  &  8.71 & 5 &  6 & 6 & 18  \\ 
 0.3   &  6.18 & 2 &  6 & 13 & 19 & 0.85 &  4.81 & 5 &  6 & 6 & 18  \\ 
 0.35  &  8.98 & 2 &  6 & 13 & 19 & 0.9  &  4.16 & 5 & 13 & 6 &  9  \\ 
 0.4   & 12.60 & 5 &  6 &  6 & 18 & 0.95 &  4.18 & 5 & 13 & 6 &  9  \\
 0.45  & 14.21 & 5 &  6 &  6 & 18 & 1.0  &  4.13 & 5 & 13 & 6 &  9  \\ 
 0.5   & 14.72 & 5 &  6 &  6 & 18 &      &       &   &    &   &
\end{tabular}
\end{center}
\caption{Inverse participation ratio and position of the maximum of
the scalar density as a function of the boundary condition parameter 
$\zeta$. The data are for configuration 125 of the confined ensemble
($6\times20^3, \beta = 8.20$).
\label{125table}}
\end{table}

\begin{figure}[t!]
\hspace*{-15mm}
\epsfig{file=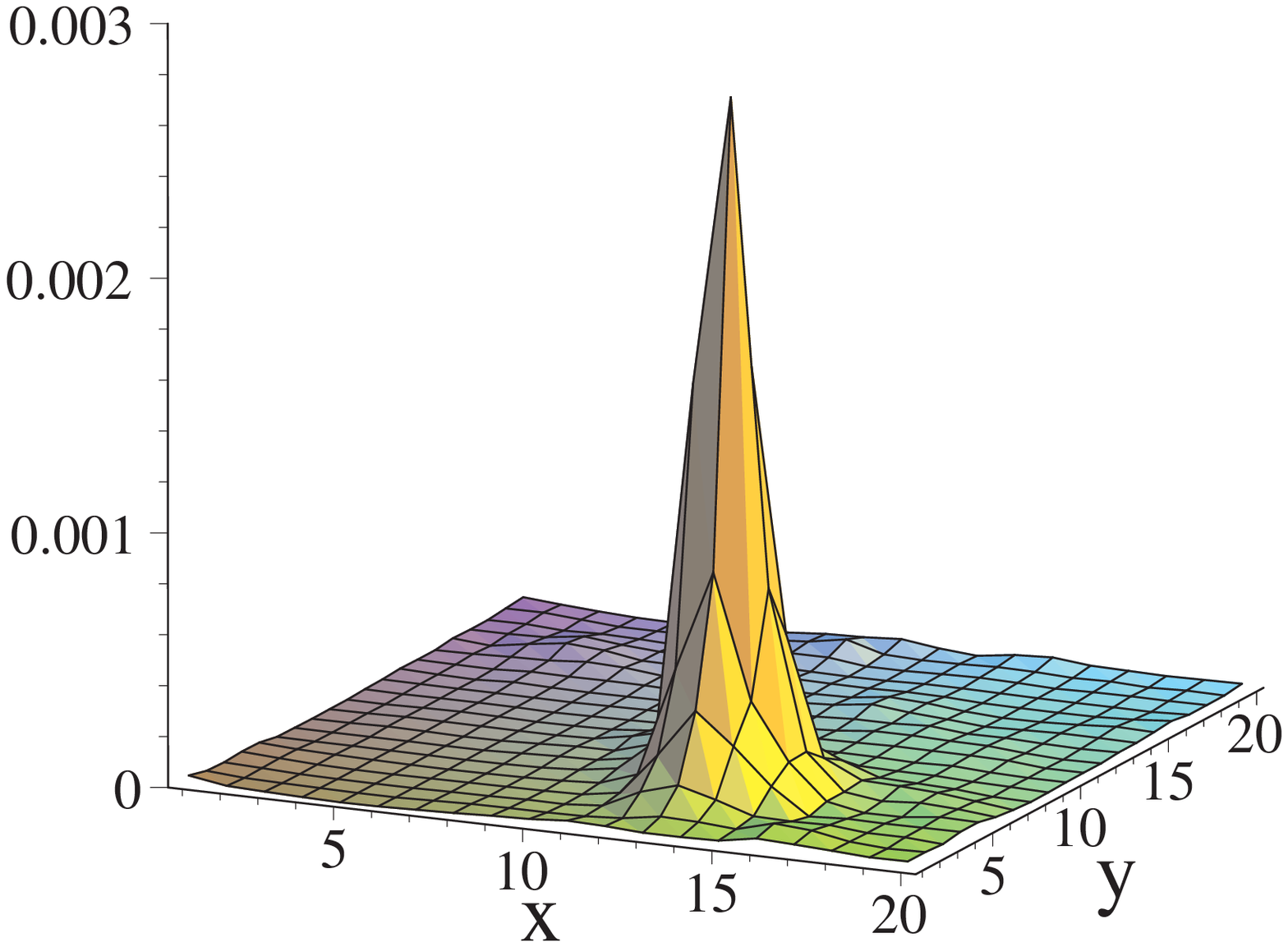,height=3.5cm,clip}
\hspace{-3mm}
\epsfig{file=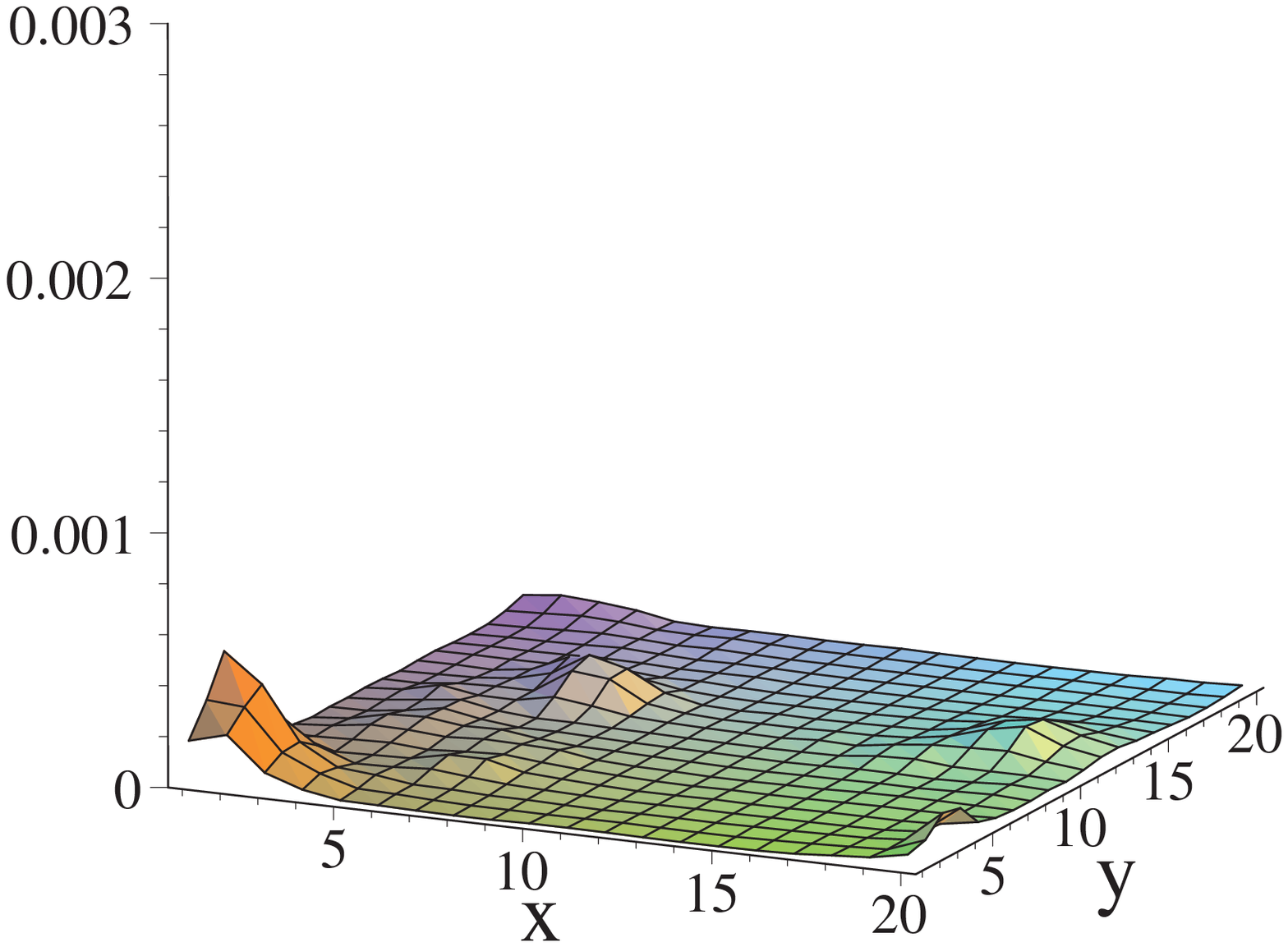,height=3.5cm,clip}
\hspace{-3mm}
\epsfig{file=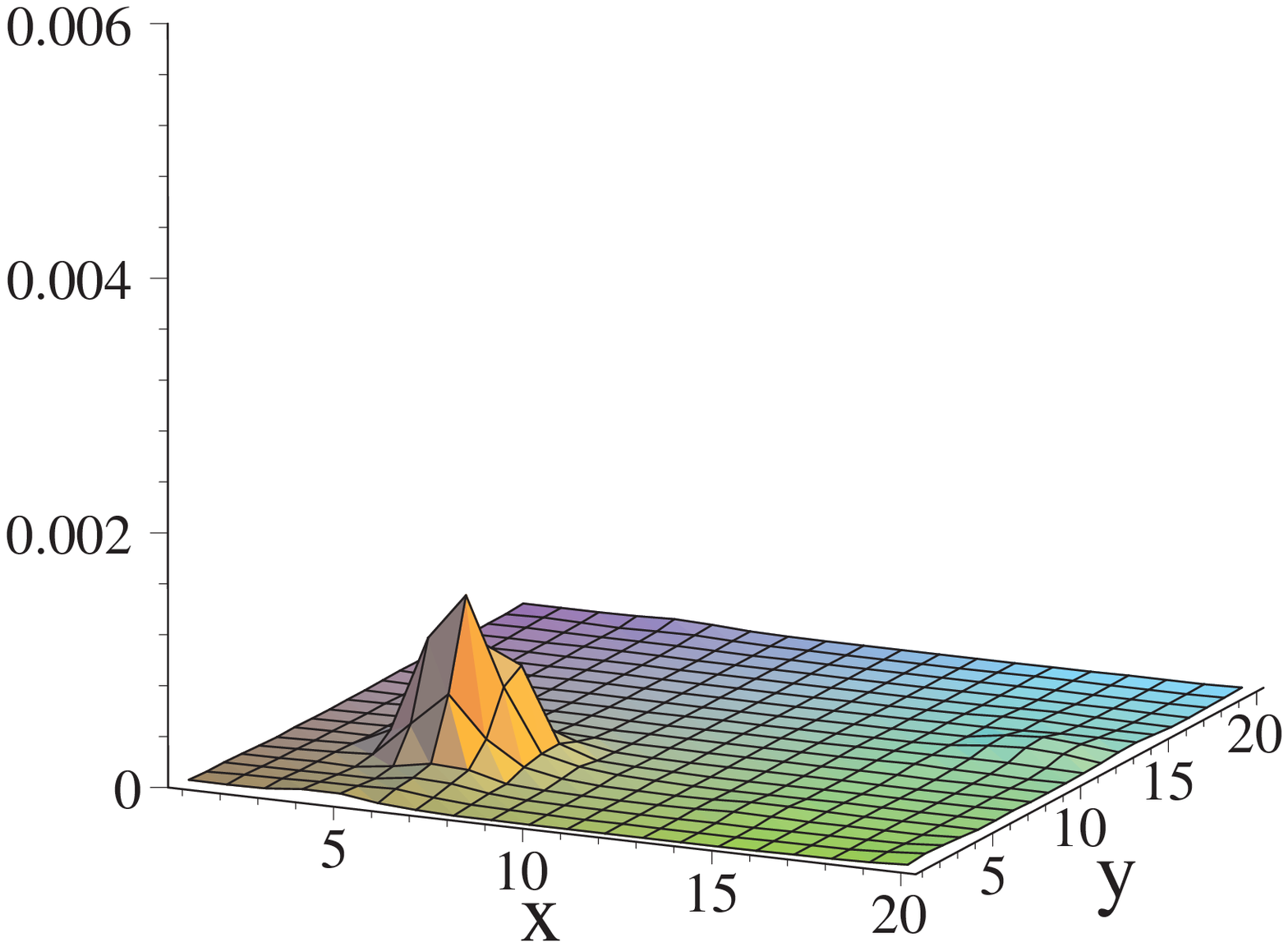,height=3.5cm,clip} 
\\
\hspace*{-15mm}
\epsfig{file=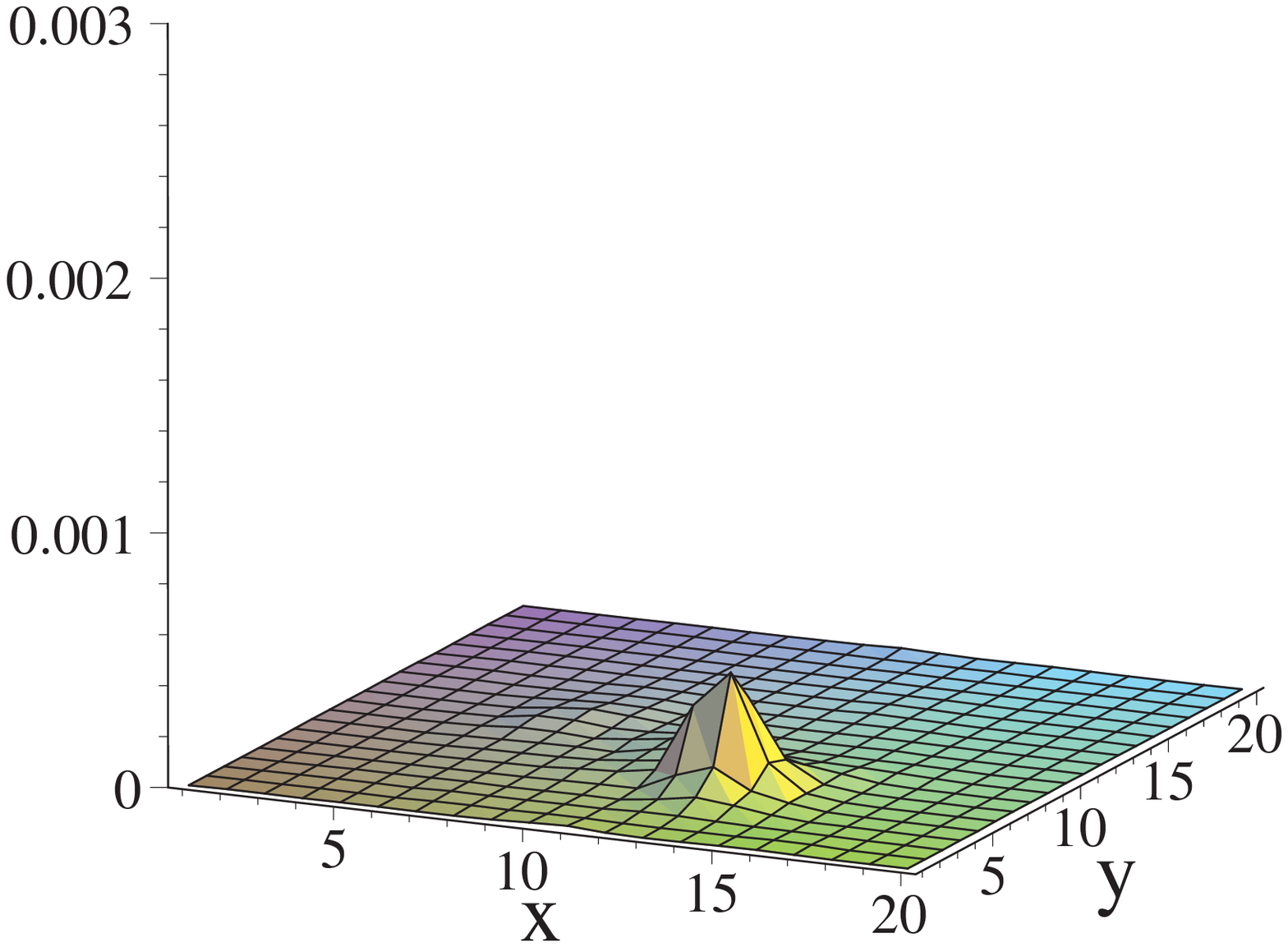,height=3.5cm,clip}
\hspace{-3mm}
\epsfig{file=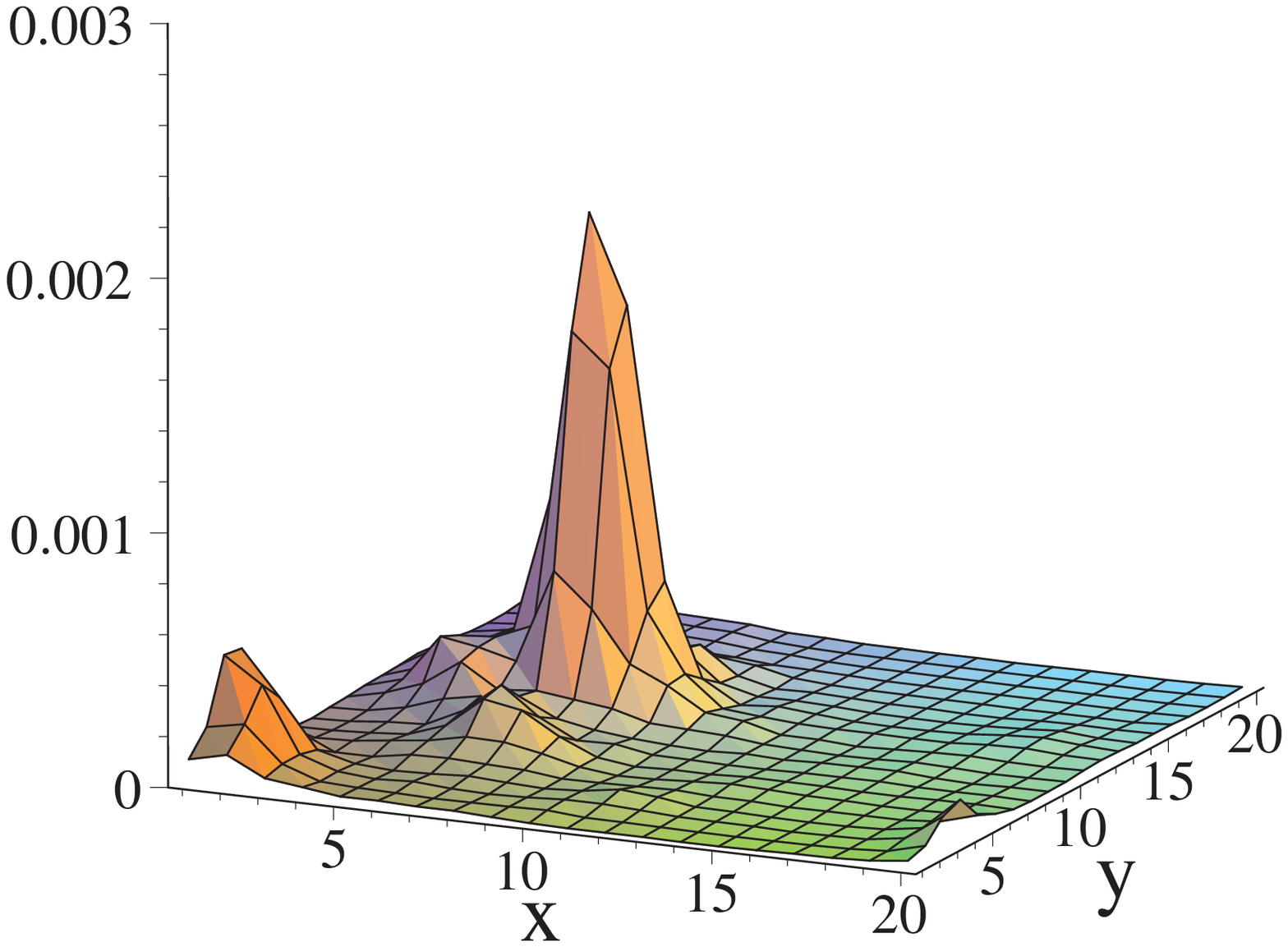,height=3.5cm,clip}
\hspace{-3mm}
\epsfig{file=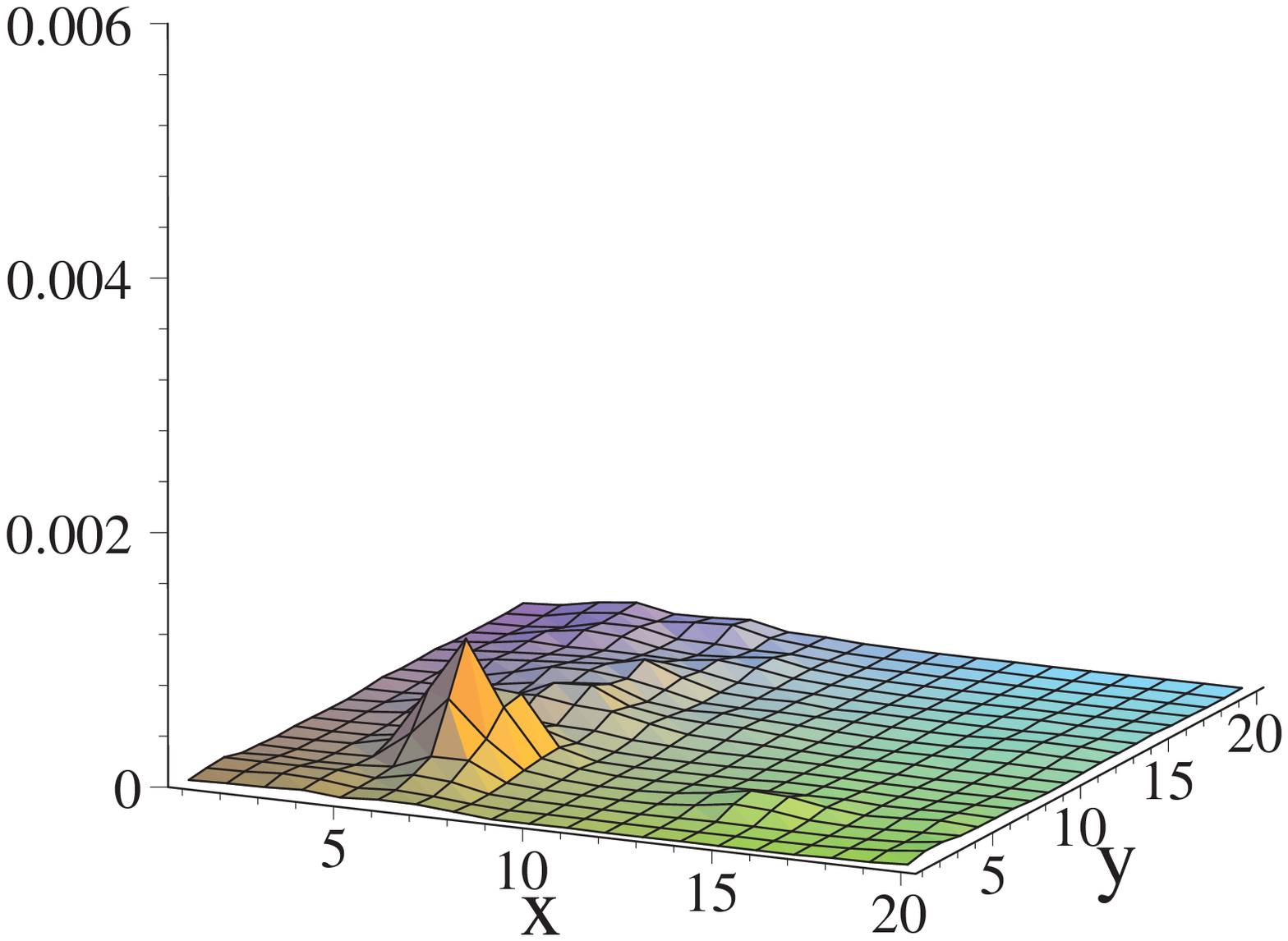,height=3.5cm,clip} 
\vspace{5mm}
\hspace*{-15mm}
\epsfig{file=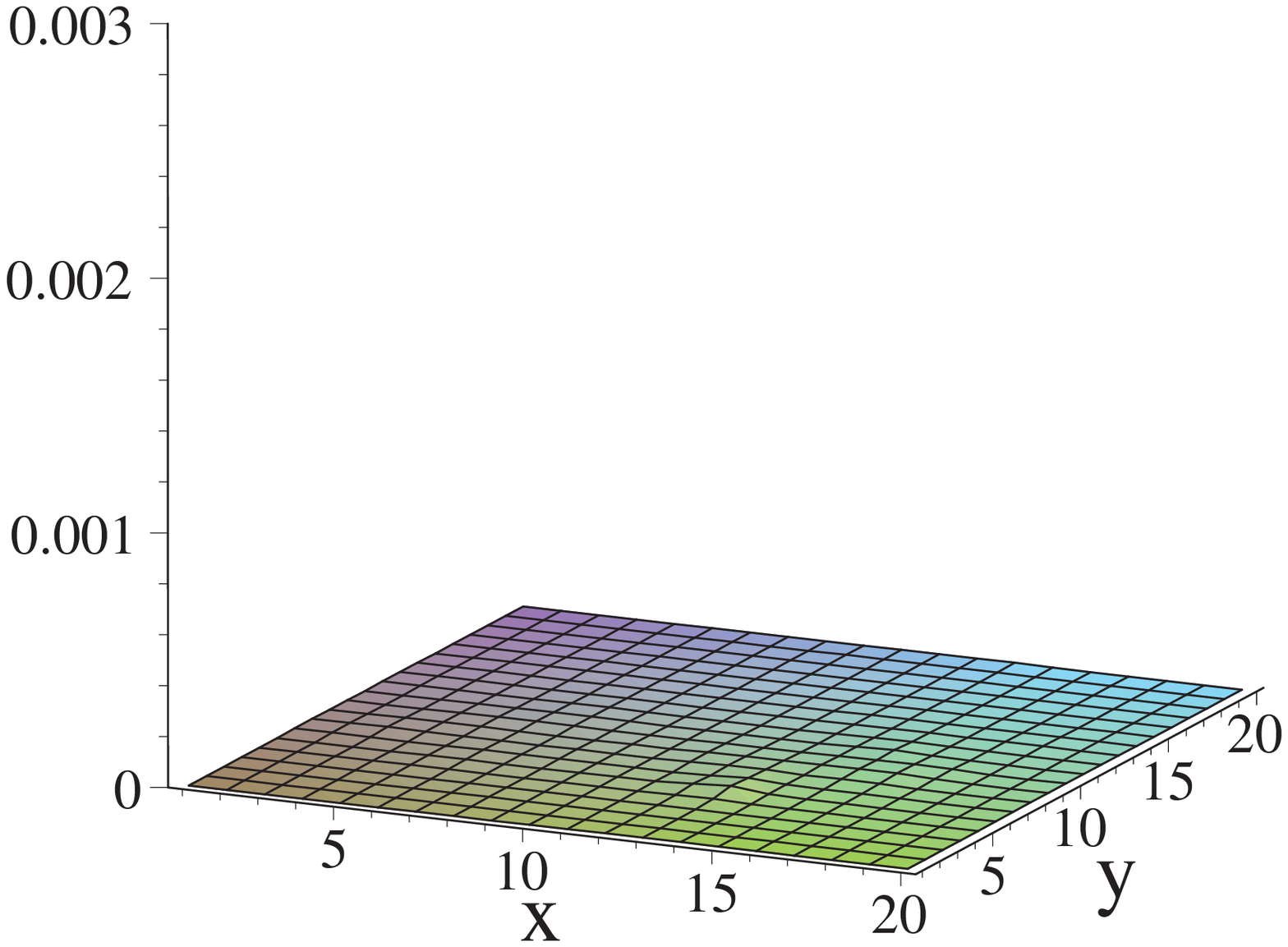,height=3.5cm,clip}
\hspace{-3mm}
\epsfig{file=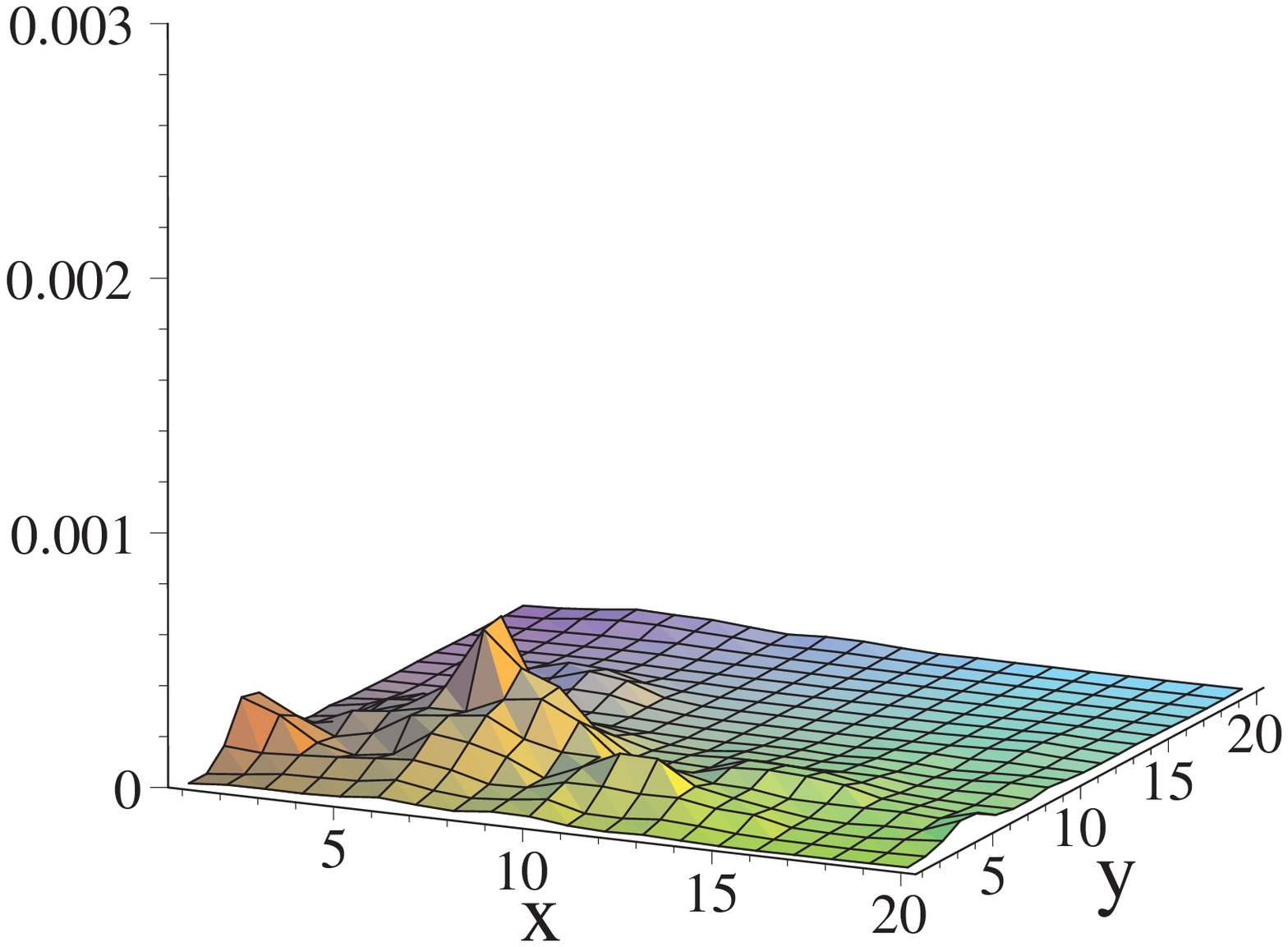,height=3.5cm,clip}
\hspace{-3mm}
\epsfig{file=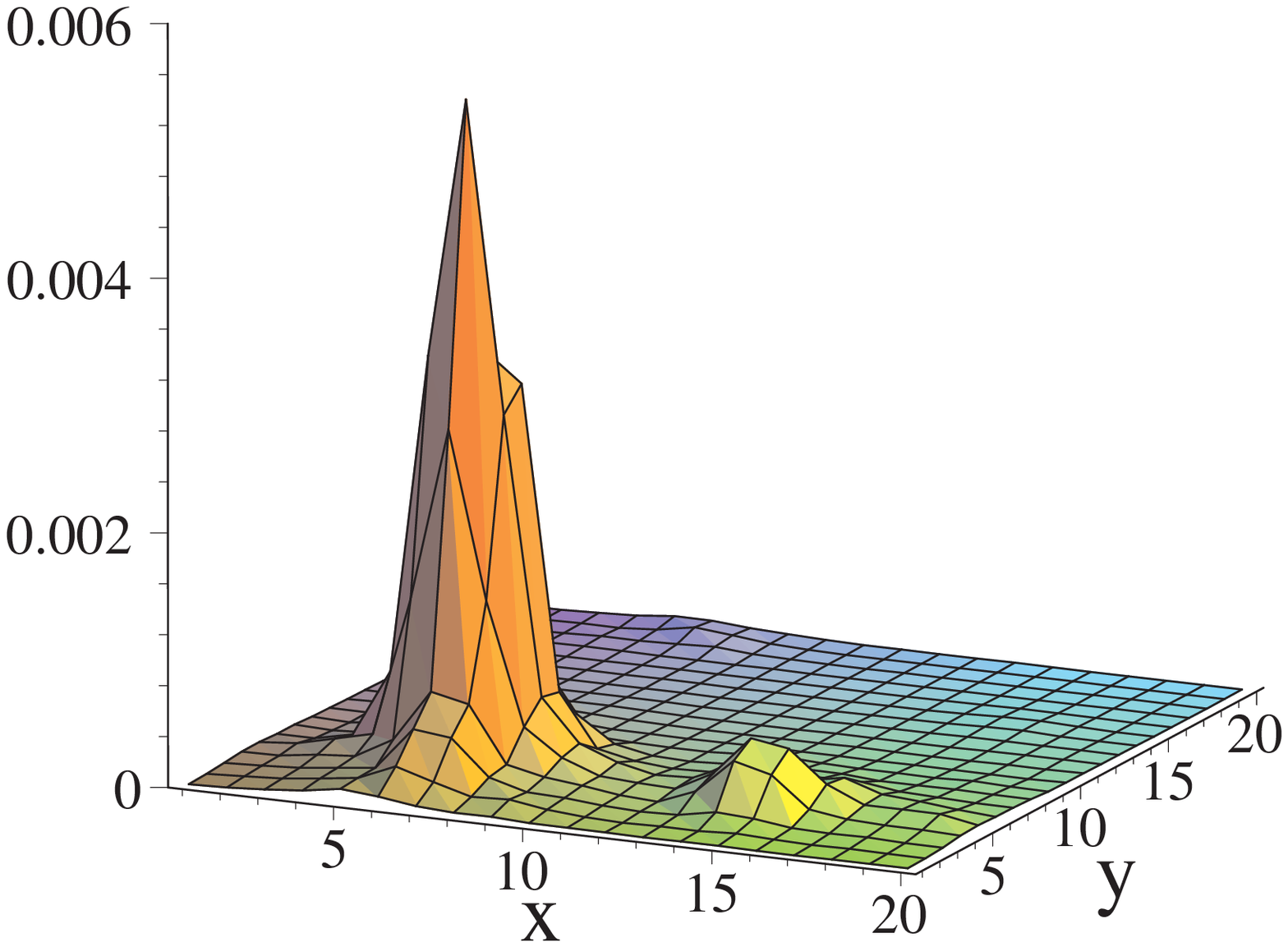,height=3.5cm,clip} 
\\
\caption{Slices of the scalar density for $6\times20^3, \beta = 8.20$,
configuration 125. 
We show $x,y$-slices at 
$t = 5$, $z = 9$ (left column), at $t = 2$, $z = 19$ (center column)
and $t = 5$, $z = 18$ (right column). The values for $\zeta$ are
$\zeta = 0.05, 0.3, 0.65$ (from top to bottom).
\label{dens20b820_125a}}
\end{figure}

Table \ref{125table} shows that the zero-mode of configuration
125 visits three different lumps 
located at $(t,x,y,z)$ = (5,13,6,9), (2,6,13,19) and (5,6,6,18).
These three lumps are well separated from each other with mutual
4-distances of $d_4 = 14.38$, $d_4 = 11.40$ and $d_4 = 7.68$ in units
of the lattice spacing. In fermi this is $d_4 =$ 1.65 fm, 1.31 fm and 0.88 fm. 
The most prominent lump is the one at $(t,x,y,z) = (5,6,6,18)$ which
is seen by the zero mode at all $\zeta$
between $\zeta = 0.4$ and $\zeta = 0.85$. 
We remark that this lump is also the most localized one, i.e.\ reaches
the largest values for the inverse participation ratio. This is as expected
for the KvB zero-modes, where the mass of the underlying monopole 
is proportional to $\mu_{m+1} - \mu_m$ (compare Eq.~\ref{monomass}).
Thus the lump which occupies the largest interval in $\zeta$ is
also the most localized one. This pattern also holds for all other 
configurations listed in Table \ref{confs20b820}. It is interesting to note 
that the lump at $(t,x,y,z) = (5,6,6,18)$ already briefly
appears at $\zeta = 0.1,0.15$ before the lump at $(t,x,y,z) = (2,6,13,19)$
takes over. The reason for that is that the lump at (2,6,13,19) starts to grow
only slowly such that for the small window in $\zeta$ the remains
of the tallest lump at (5,6,6,18) dominate.

Let us now look at plots for the scalar density for this configuration. 
In Fig.\ \ref{dens20b820_125a} we show 
$x,y$-slices for the scalar density taken through the maxima of
the three lumps, i.e.\ at $t = 5$, $z = 9$ (left column), 
at $t = 2$, $z = 19$ (center column)
and $t = 5$, $z = 18$ (right column). The values for $\zeta$ are
$\zeta = 0.05, 0.3, 0.65$. These are the values where the dominating lumps
have their maximum. In between these extremal values one has intermediate 
situations with a receding and an advancing lump. Note
that the plots in the r.h.s.\ column have a different scale
and the corresponding lump is about twice as tall as the other two lumps.

It is interesting to inspect also other slices, in particular 
a slice containing the time direction. For an unperturbed KvB solution
one expects that the lump extends over all of the time direction 
and that the height of the lump displays an oscillating 
behavior in time. For the action 
density of cooled SU(2) lattice configurations with twisted boundary
conditions such a behavior was observed in \cite{garciaetal99}. 
 
\begin{figure}[t]
\hspace*{-7mm}
\epsfig{file=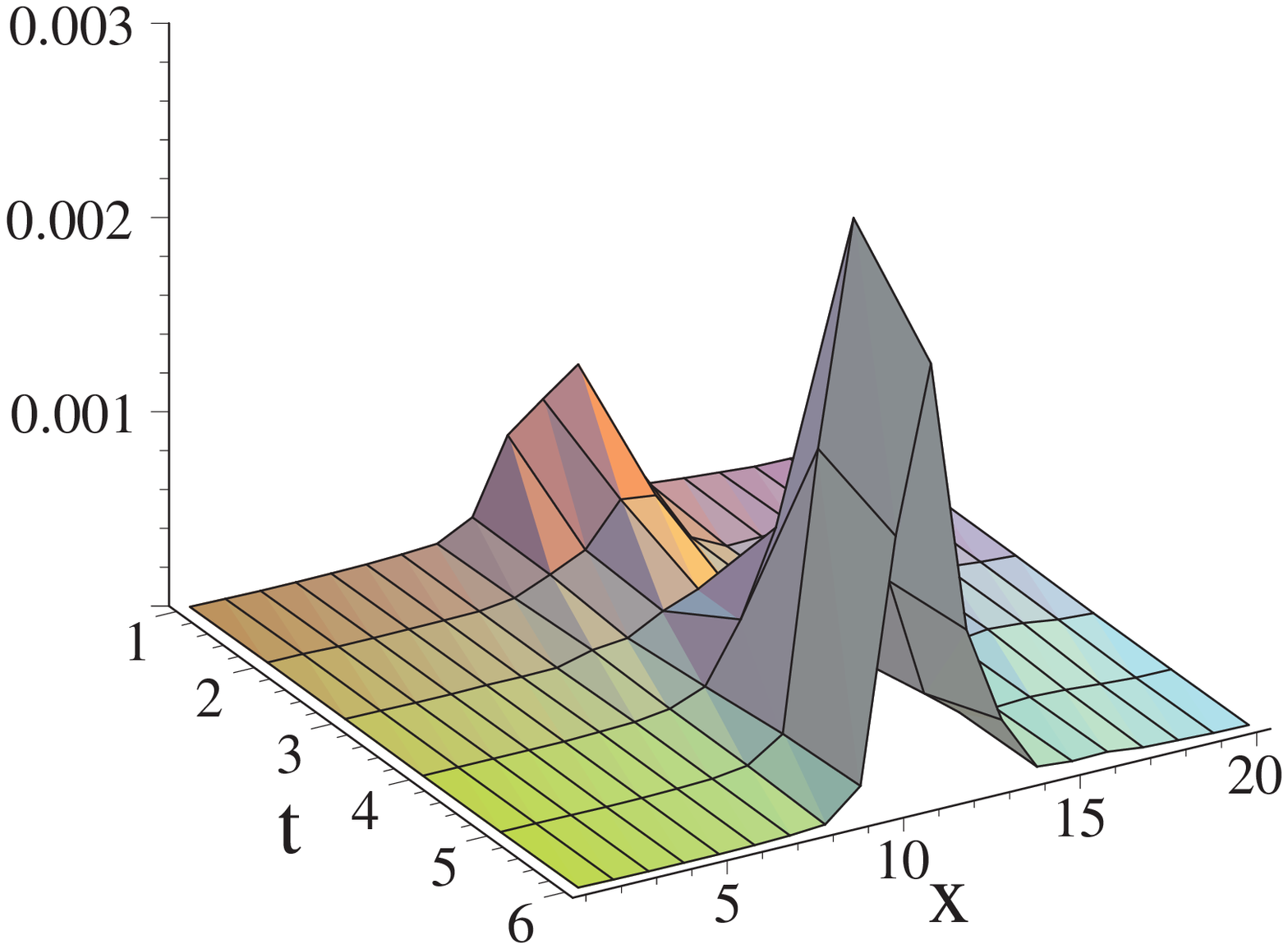,height=3.5cm,clip}
\hspace{-6mm}
\epsfig{file=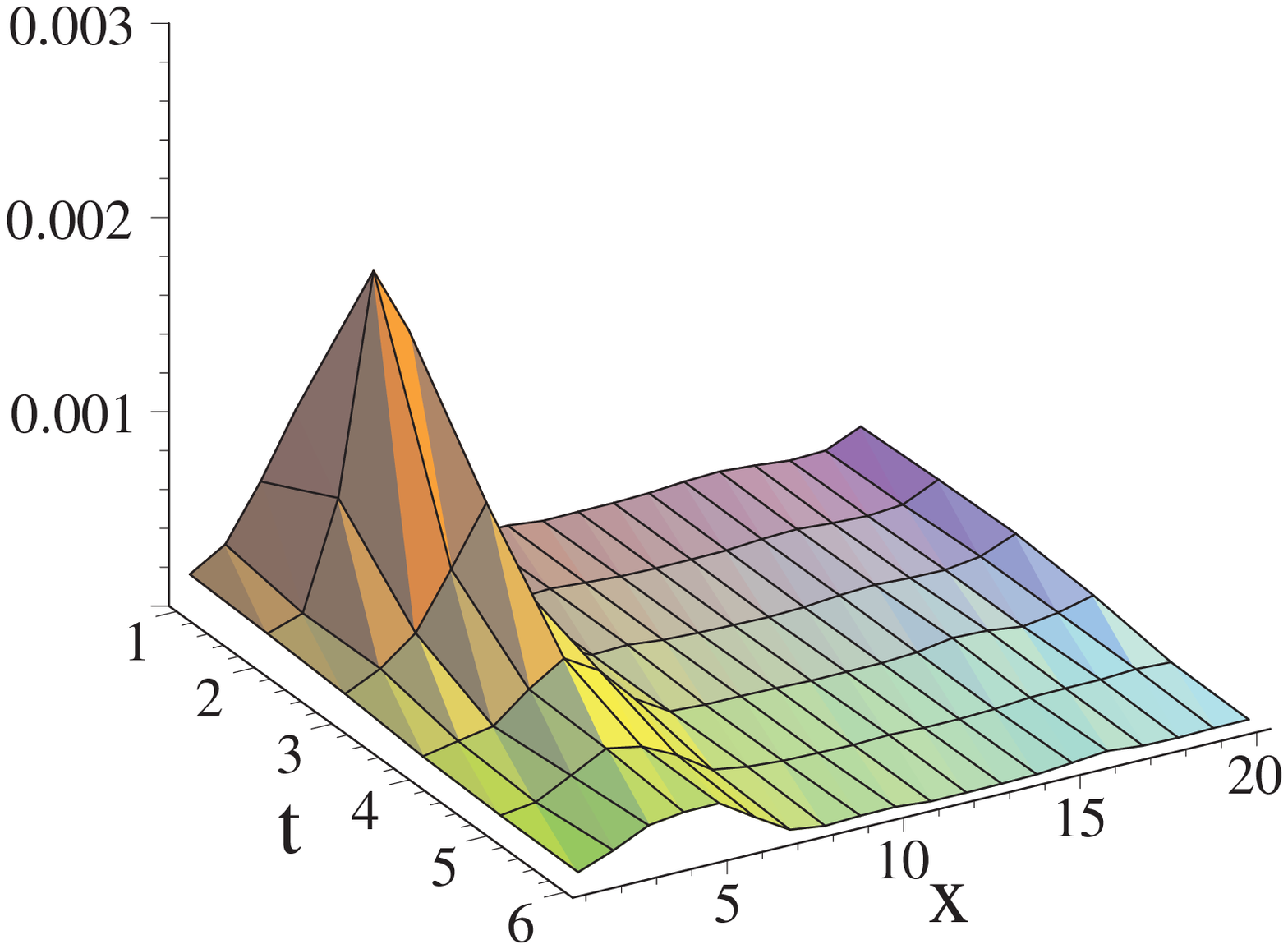,height=3.5cm,clip}
\hspace{-6mm}
\epsfig{file=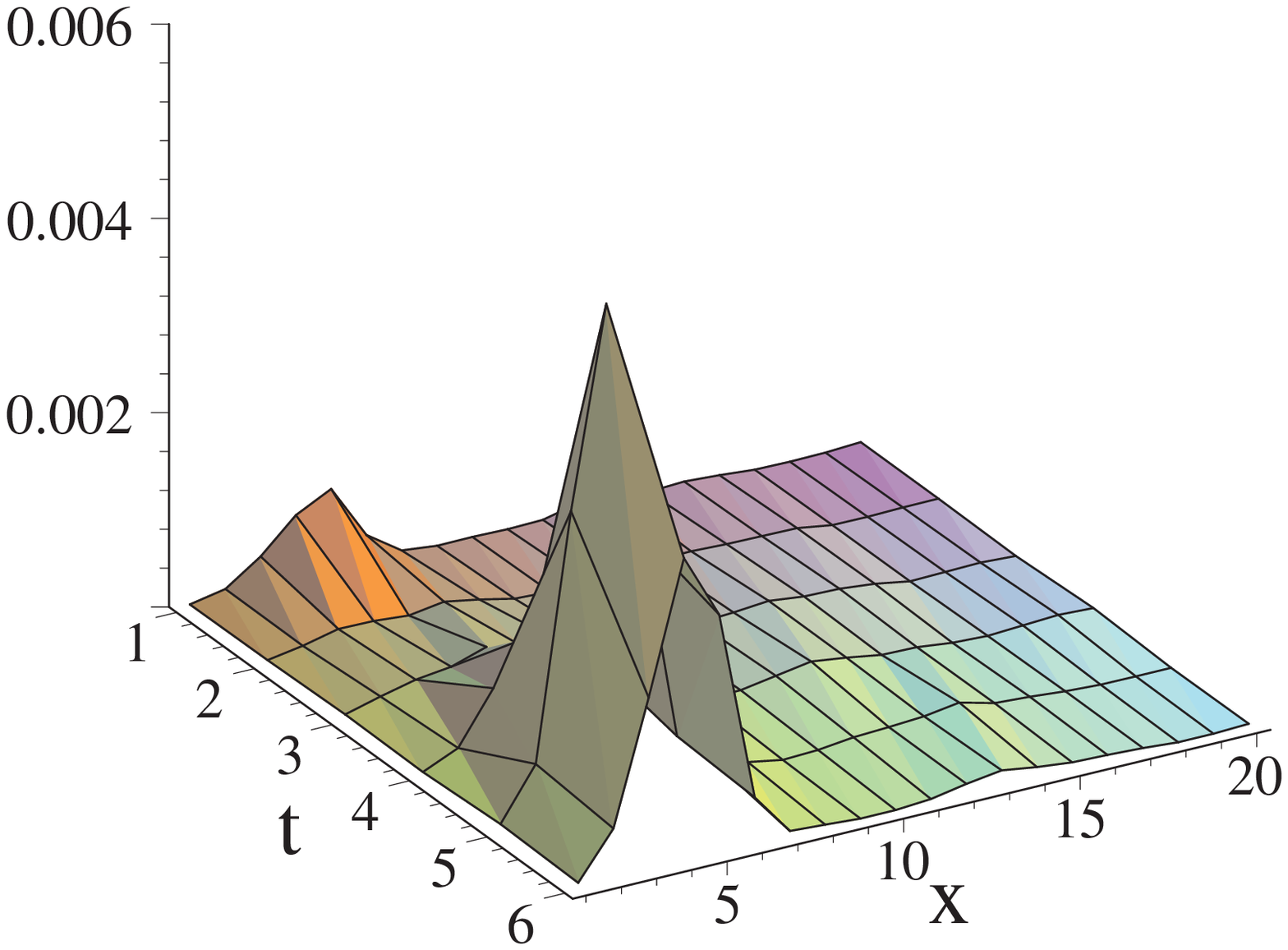,height=3.5cm,clip} 
\\
\caption{Slices of the scalar density for $6\times20^3, \beta = 8.20$,
configuration 125. 
We show $t,x$-slices at 
$y = 6$, $z = 9$ for $\zeta = 0.05$ (left plot), 
at $y = 13$, $z = 19$ for $\zeta = 0.3$ (center plot)
and $y = 6$, $z = 18$ for $\zeta = 0.65$ (right plot). 
These plots correspond to the three dominant lumps
also shown in Figs.\ \protect{\ref{dens20b820_125a}}.
\label{dens20b820_125c}}
\end{figure}

In Fig.\ \ref{dens20b820_125c} we show $t,x$-slices of the scalar density 
again for configuration 125 which was already used for the plots in 
Fig.\ \ref{dens20b820_125a}. In particular we 
show the slices through the 3 lumps at their dominant value of $\zeta$.
They were taken at $y = 6$, $z = 9$ for $\zeta = 0.05$ (left column), 
at $y = 13$, $z = 19$ for $\zeta = 0.3$ (center column)
and $y = 6$, $z = 18$ for $\zeta = 0.65$ (right column). 
The plots show that
all three lumps are stretched along the $t$-axis. A similar behavior was 
also observed for other configurations, both below and above $T_c$.

\subsection{Results from the whole sample}

As for the deconfined phase, also for the ensemble below $T_c$ we now 
study observables evaluated for the whole ensemble in order to
establish properties of the zero-modes beyond a demonstration in
a single example. 

\begin{figure}[t]
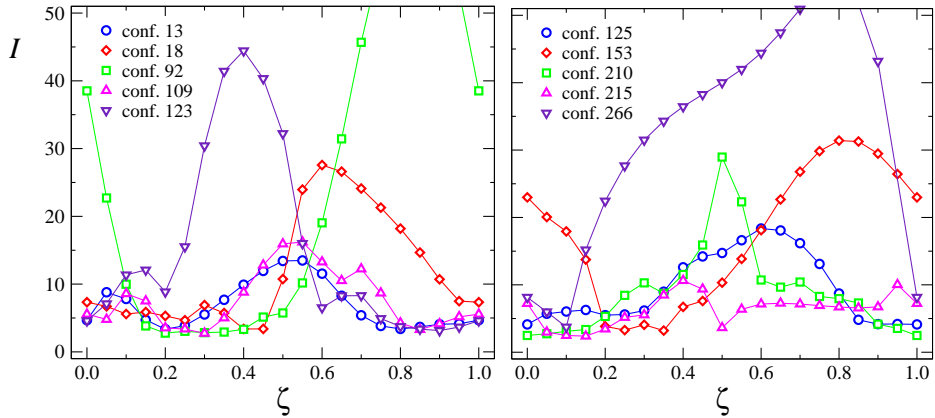

\vspace{5mm}
\begin{center}
\hspace*{-3mm}
\epsfig{file=ipr_vs_phase_20_b820_a.eps,height=5.5cm,clip}
\hspace{-1mm}
\epsfig{file=ipr_vs_phase_20_b820_b.eps,height=5.5cm,clip}
\end{center}
\vspace{-3mm}
\caption{Inverse participation ratio $I$ as a function of $\zeta$ for
the $6\times20^3$, $\beta = 8.20$ ensemble. In order to avoid overcrowded 
figures we show the data for configurations 13, 18, 92, 109 and 123 on the 
l.h.s., and configurations 125, 153, 210, 215, 266 on the r.h.s.
\label{ipr_vs_phase_20_b820}}
\end{figure}

We start with plots for the inverse participation ratio as a function
of $\zeta$. In Fig.\ \ref{ipr_vs_phase_20_b820} we show such figures for
the 10 configurations with finely spaced $\zeta$ listed in Table 
\ref{confs20b820}. To avoid overcrowding of the plots
we split the figure into two 
plots and show the data for configurations 13, 18, 92, 109 and 123 on the 
l.h.s., and configurations 125, 153, 210, 215, 266 on the r.h.s. 

From the plots it is obvious that the localization of the zero-mode
fluctuates strongly during a complete cycle through the boundary condition. 
Typically there is a single pronounced maximum in the inverse participation
ratio and one or two local maxima. The tallest peak occupies
also the largest interval of $\zeta$ in accordance with the mass formula
Eq.~(\ref{monomass}).
On the other hand it is obvious from the plots that the minima are not
necessarily located at exactly $\zeta = 0$, $\zeta = 1/3$ and 
$\zeta = 2/3$ as would be the case for an unperturbed KvB zero-mode. 
However, one should not forget that our configurations are taken 
from a thermalized ensemble containing all quantum effects. It is 
not yet known \cite{private}
in which way quantum effects modify the classical KvB 
solutions.

\begin{figure}[t!]
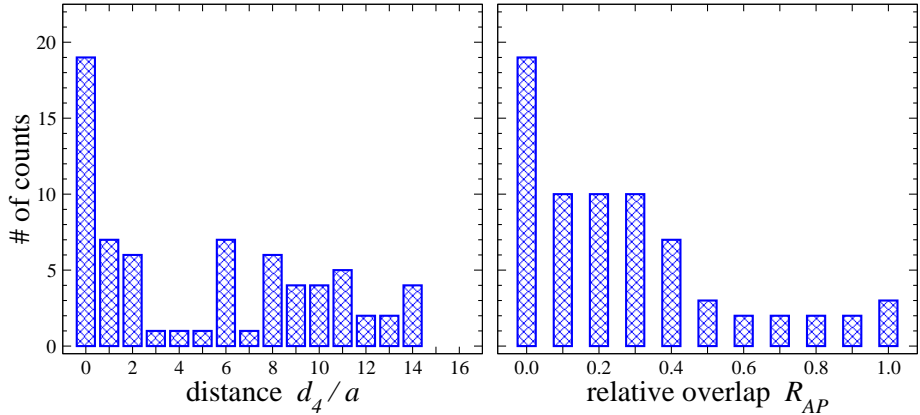

\begin{center}
\vspace{5mm}
\hspace*{-5mm}
\epsfig{file=disthisto6x20b820.eps,height=5.5cm,clip}
\epsfig{file=overlaphisto6x20b820.eps,height=5.5cm,clip}
\end{center}
\vspace{-3mm}
\caption{Histograms for the distance $d_4$ 
between the peaks in the scalar density 
when comparing periodic to anti-periodic b.c.\ (l.h.s.\ plot) 
and for the relative overlap $R_{AP}$ between 
the two zero-modes (r.h.s.\ plot). The data are for the 
ensemble in the confining phase on $6\times20^3$ lattices at $\beta = 8.20$.
\label{disthisto20b820}}
\end{figure}

In Fig.\ \ref{disthisto20b820} we show histograms for the distance 
$d_4$ between 
the lump in the periodic and anti-periodic zero-modes in the l.h.s.\ plot.
On the r.h.s.\ we show a histogram for the relative overlap $R_{AP}$
between these 
two lumps. These two histograms contain the data for all 70 configurations
in the ensemble. 

The histogram on the l.h.s.\ shows that for half of the configurations 
the zero-mode for periodic boundary conditions and its anti-periodic 
counterpart sit on lumps which are more than 5 lattice spacings apart,
i.e.\ at least 0.69 fm. The plot on the r.h.s.\ supports this finding
by illustrating that typically the two lumps do not overlap 
substantially. Note that even if the lumps are located at the same position 
their relative overlap can be smaller than 1 when they differ in size.
Both plots in Fig.\ \ref{disthisto20b820} show that configurations 
where the zero-mode changes its location and size are not singular
events but make up a large portion of the ensemble.

\section{Results on the torus}

\subsection{Overview over the configurations}

Having successfully identified the substructure predicted for the KvB 
zero-modes in ensembles with high temperature it is natural to 
apply the same techniques to configurations at low temperature. 
In this section we perform such an exploratory study and 
analyze our ensemble on $16^4$ lattices with 
$\beta = 8.45$. The difference to the high temperature configurations
is that the lattice geometry does not single out a time direction. 
We break the euclidean symmetry explicitly and use our generalized 
fermionic boundary condition for the 1-direction, while for the other 
directions we implement periodic boundary conditions. We remark that 
on the torus no analytic results are available to compare with. 

As for the cases with temperature we compute eigenvectors of the Dirac operator
with different values of the boundary condition parameter $\zeta$.
For all 46 configurations of the ensemble (compare Table \ref{confdata}) 
we use $\zeta = 0$ and $\zeta = 0.5$, i.e.~periodic and anti-periodic 
boundary conditions. For a subensemble of 10 configurations we 
computed eigenvectors for all values of $\zeta$ between 0 and 1 in steps of
0.1.

\begin{table}[b!]
\begin{center}
\hspace*{-5mm}
\begin{tabular}{r|r|r|cccc|cccc|r|c}
conf. & $I_{P} $ & $I_{A}$ 
& 
\multicolumn{4}{c}{$(t,x,y,z)^{max}_{P}$} \vline & 
\multicolumn{4}{c}{$(t,x,y,z)^{max}_{A}$} \vline & $d_4/a$ & \# lumps\\
\hline
32  &  6.84 &   8.62 &  5 &  9 &  5 &  8 & 15 &  5 &  8 & 15 & 10.48 & 2 \\
61  & 51.85 &  66.22 & 14 & 11 &  8 &  7 & 14 & 11 &  8 &  7 &  0.00 & 1 \\
64  &  3.08 &   3.43 & 11 & 12 & 13 &  2 &  7 &  2 & 12 &  5 &  7.87 & 3 \\
82  &  3.96 &   3.98 &  3 & 15 &  5 &  5 &  4 &  1 & 15 &  2 &  7.07 & 2 \\
93  &  2.47 &   3.00 &  3 &  7 &  5 & 14 &  8 & 11 &  4 &  6 & 10.29 & 3 \\
103 &  2.54 &   3.04 &  4 & 14 &  3 &  2 &  4 & 14 &  3 &  2 &  0.00 & 1 \\
351 &  3.06 &   3.73 & 14 & 10 &  3 &  7 &  7 &  4 &  6 & 12 & 10.90 & 3 \\
354 &  7.67 &  11.86 & 13 & 14 &  4 &  4 & 13 & 14 &  4 &  4 &  0.00 & 1 \\
522 &  7.32 &   2.71 &  3 & 14 & 16 &  6 & 16 & 11 & 15 &  7 &  4.47 & 3 \\
561 &  3.94 &   6.74 &  8 &  3 & 15 &  8 & 14 &  4 &  9 &  9 &  8.60 & 1 
\end{tabular}
\end{center}
\caption{Parameters of our ensemble on the torus ($16^4, \beta = 8.45$).
We list the configuration number,
the inverse participation ratios 
$I_{P}$ and $I_{A}$ for the zero-modes with periodic, 
respectively anti-periodic b.c., the position of the maxima for
the corresponding zero-modes, the 4-distance $d_4$ between these two 
maxima and the number of different lumps visited by the mode 
in a complete cycle of $\zeta$.
\label{confsz16b845}}
\end{table}

For these 10 configurations Table \ref{confsz16b845} 
shows some basic properties.
In particular we list the configuration number, the inverse participation 
ratio for the periodic and anti-periodic zero-modes, the  
positions of the maxima in the corresponding scalar densities and the difference
between these two peaks. In addition we display the number of 
lumps visited by the zero-mode in a complete cycle through the phase at the 
boundary. Here we count a lump as independent if it appears at 
two subsequent values
of $\zeta$, i.e.~is visible in a $\zeta$-interval of at least length 0.1.
This is the same criterion used for the $6\times20^3, \beta = 8.20$ ensemble.

The table makes obvious that also for the ensemble on the torus we do find
configurations where the zero-mode is located at different positions
for different values of $\zeta$. The 4-distances between the periodic and 
the anti-periodic lumps reach values up to 11 in lattice units which is 
up to 1.03 fm. When discussing the complete ensemble we will find that about
half of the configurations have a sizeable distance between the two peaks. 
The zero-modes visit between 1 and 3 different lumps in a complete cycle 
of $\zeta$. We remark that although for configuration 561 we have 
$d_4 \neq 0$ we list only a single lump in the last column. The reason is that 
only for $\zeta = 0$ the zero-mode is located at a different position and
thus the criterion of two subsequent appearances is not fulfilled. The 
pattern observed for this configuration is similar to the pattern
of a configuration in the deconfined phase with real Polyakov loop.

\subsection{Example for the generic behavior}

Let us now look at figures of the scalar density for a typical configuration. 
In particular we will show plots for configuration 32 of our ensemble.
Before presenting these figures we discuss Table \ref{32table} where we 
list some properties of this configuration as a function of $\zeta$.
In the left half of the table we show $\zeta$, the inverse participation 
ratio and the position of the maximum. The zero-mode changes its position at 
$\zeta = 0.3$ and $\zeta = 0.9$. The localization shows some mild
changes with reaching local maxima at $\zeta = 0.5$ and $\zeta = 0.1$,
i.e.~approximately in the middle of the $\zeta$-intervals for the two lumps.

In the right half of the table we show the data from a small consistency
check which we performed for 4 of the 10 configurations. We switched 
the direction where we apply the boundary condition from the 1-direction
to the 2-direction. Since our lattice does not single out a particular 
direction any direction for applying the boundary condition serves 
equally well and the results should be consistent. The data in the right 
half of the table show that this is indeed the case. Also with the 
non-trivial boundary condition 
applied in the 2-direction we see the same two lumps 
at (5,9,5,8) and (15,5,8,15). They appear, however, at different values 
of $\zeta$ which is not surprising since a single gauge configuration 
does not respect rotational symmetry. Also for the other 3 configurations
where we experimented with changing the direction we single out
by the boundary condition, we find that essentially
the same lumps are seen, again at slightly different values of $\zeta$.  

\begin{table}[t!]
\begin{center}
\hspace*{-5mm}
\begin{tabular}{l|c|cccc|l|c|cccc}
\multicolumn{6}{c}{{\bf b.c.~in 1-direction:}} \vline& 
\multicolumn{6}{c}{{\bf b.c.~in 2-direction:}} \\ 
\hline
$\;\;\zeta$ & $I$ & 
\multicolumn{4}{c}{$(t,x,y,z)^{max}$} \vline & $\;\;\zeta$ & $I$ &
\multicolumn{4}{c}{$(t,x,y,z)^{max}$}\\
\hline
 0.0 & 6.84 &  5 & 9 & 5 &  8 & 0.0 & 6.84 &  5 & 9 & 5 &  8  \\
 0.1 & 7.47 &  5 & 9 & 5 &  8 & 0.1 & 6.01 &  5 & 9 & 5 &  8  \\  
 0.2 & 7.09 &  5 & 9 & 5 &  8 & 0.2 & 4.54 & 15 & 5 & 8 & 15  \\
 0.3 & 6.86 & 15 & 5 & 8 & 15 & 0.3 & 9.76 & 15 & 5 & 8 & 15  \\ 
 0.4 & 7.56 & 15 & 5 & 8 & 15 & 0.4 & 8.44 & 15 & 5 & 8 & 15  \\
 0.5 & 8.62 & 15 & 5 & 8 & 15 & 0.5 & 5.73 & 15 & 5 & 8 & 15  \\ 
 0.6 & 7.50 & 15 & 5 & 8 & 15 & 0.6 & 4.97 & 15 & 5 & 8 & 15  \\ 
 0.7 & 5.22 & 15 & 5 & 8 & 15 & 0.7 & 5.11 &  5 & 9 & 5 &  8  \\ 
 0.8 & 3.91 & 15 & 5 & 8 & 15 & 0.8 & 5.71 &  5 & 9 & 5 &  8  \\
 0.9 & 4.87 &  5 & 9 & 5 &  8 & 0.9 & 6.49 &  5 & 9 & 5 &  8  \\ 
 1.0 & 6.84 &  5 & 9 & 5 &  8 & 1.0 & 6.84 &  5 & 9 & 5 &  8
\end{tabular}
\end{center}
\caption{Inverse participation ratio and position of the maximum of
the scalar density as a function of the boundary condition parameter 
$\zeta$. The data are for configuration 32 of the ensemble on
the torus ($16^4, \beta = 8.45$). The data in the left half of the 
table were obtained with using the nontrivial boundary condition
in 1-direction, while the data in the right half is for 2-direction.
The corresponding plots of the scalar density are shown in Fig.\
\protect{\ref{densz16b845_032a}}.
\label{32table}}
\end{table}

\begin{figure}[t!]
\begin{center}
\epsfig{file=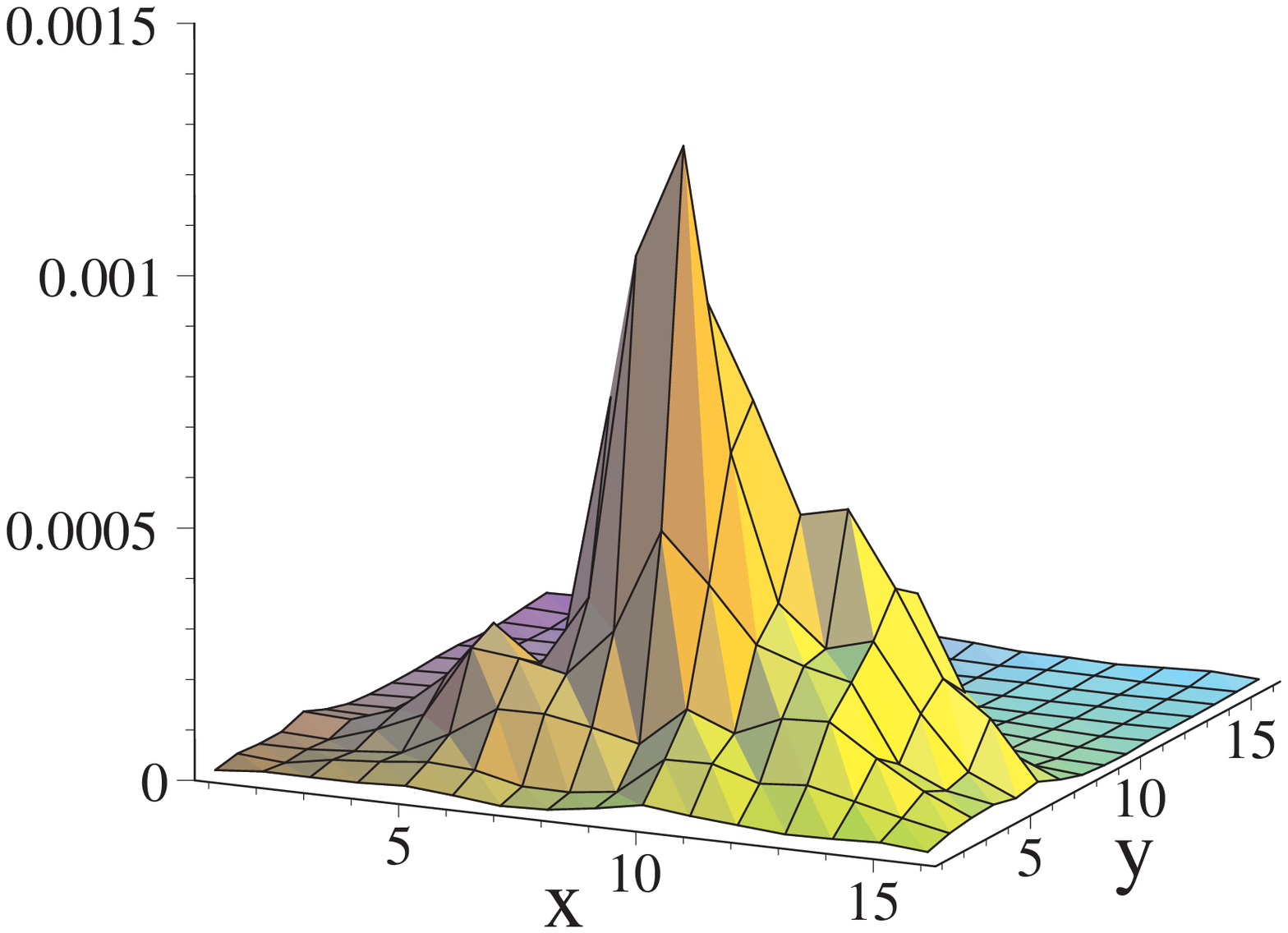,height=3.5cm,clip}
\epsfig{file=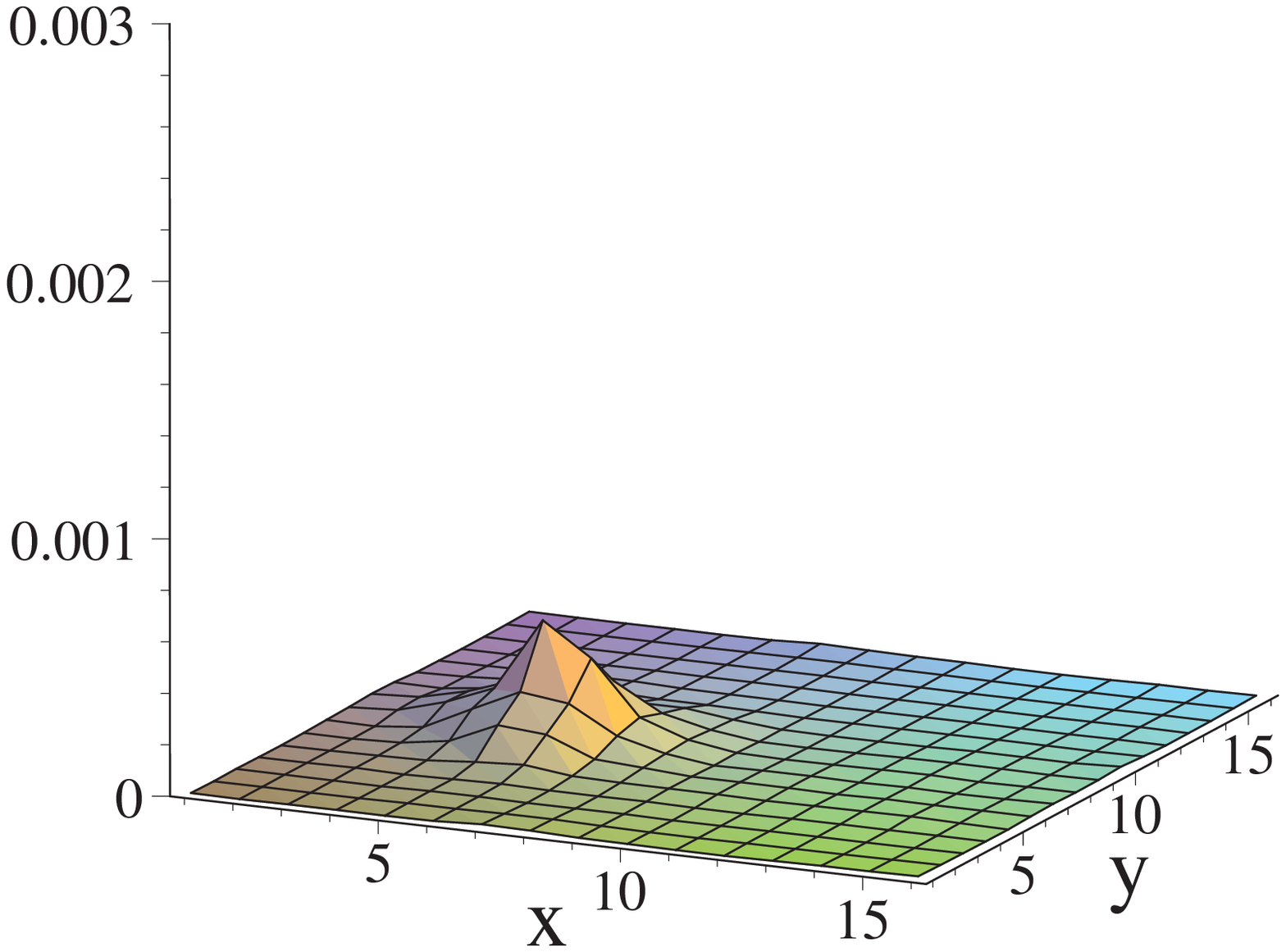,height=3.5cm,clip}
\\
\epsfig{file=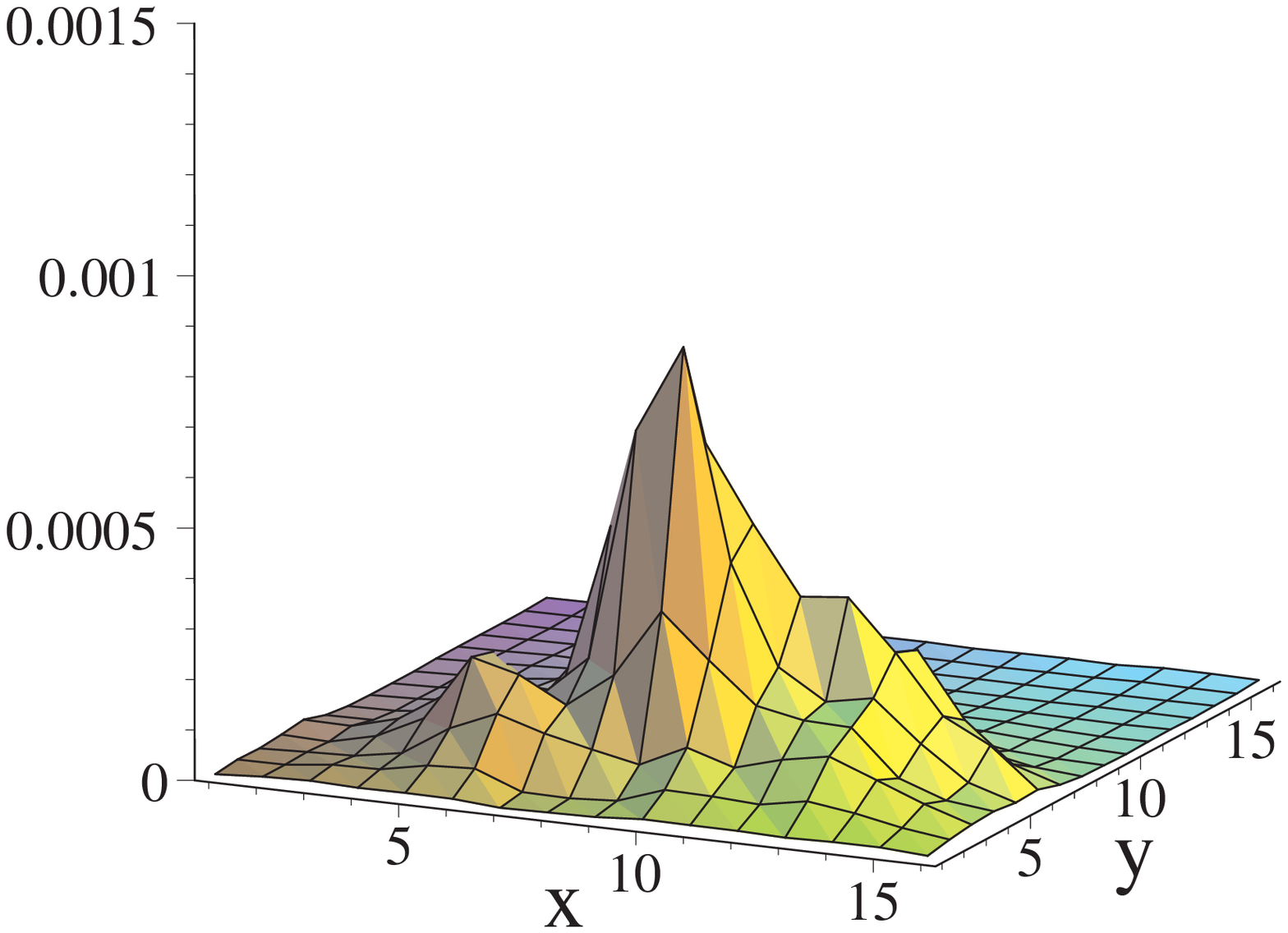,height=3.5cm,clip}
\epsfig{file=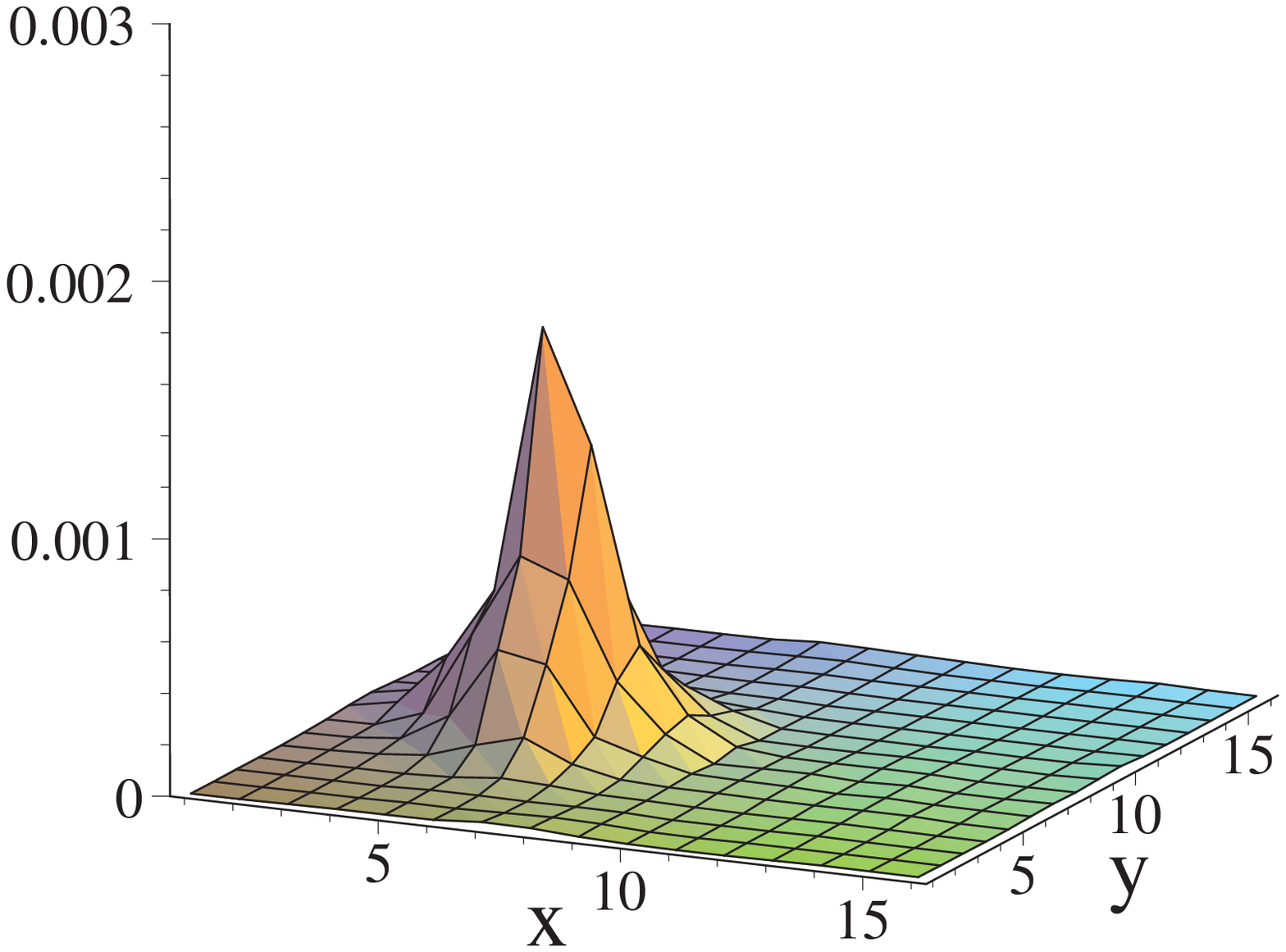,height=3.5cm,clip}
\\
\epsfig{file=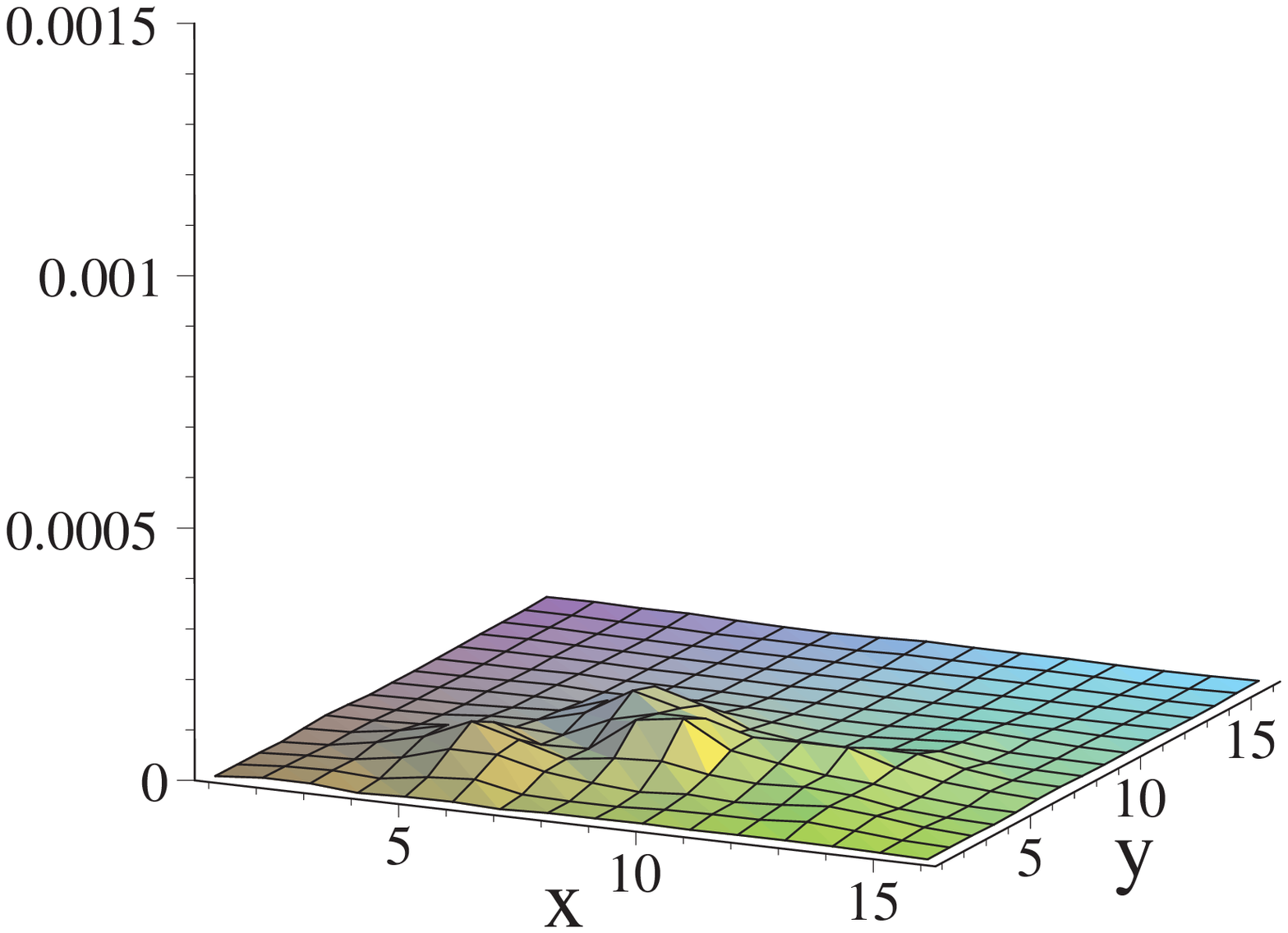,height=3.5cm,clip}
\epsfig{file=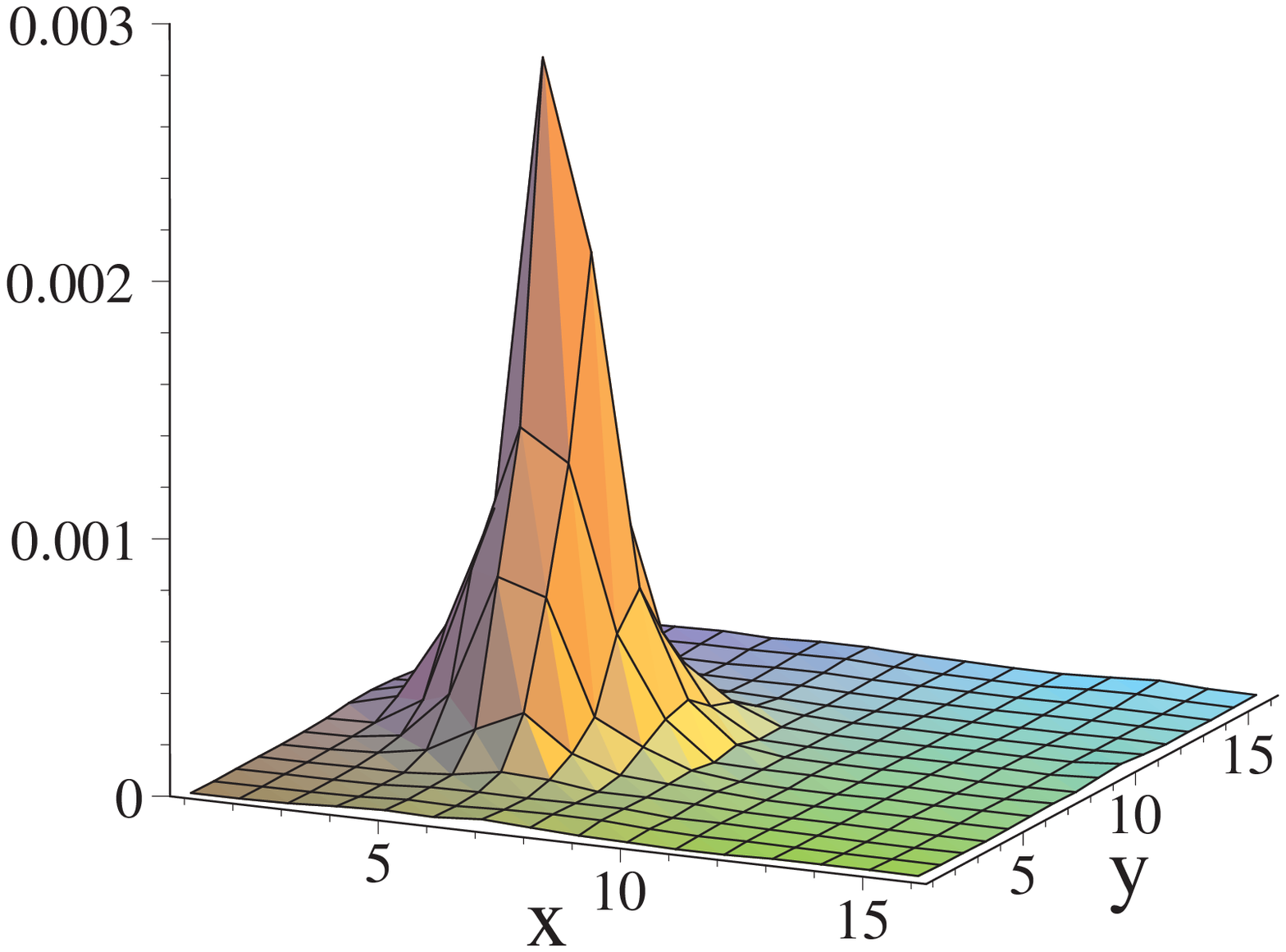,height=3.5cm,clip}
\end{center}
\vspace{5mm}
\caption{Slices of the scalar density for $16^4, 
\beta=8.45$, configuration 32. 
We show $x_2,x_3$-slices (in the plots we again 
label these two axes as $x$ and $y$) at 
$x_1 = 5$, $x_4 = 8$ (left column)  and at 
$x_1 = 15$, $x_4 = 15$ (right column). The values for $\zeta$ are
(from top to bottom) $\zeta = 0, 0.3, 0.5$. Note the 
different scale for the l.h.s.~and r.h.s.~plots.
\label{densz16b845_032a}}
\vspace{3mm}
\end{figure}

In Fig.\ \ref{densz16b845_032a} we show 
$x_2,x_3$-slices of the scalar density for configuration 32
(in the plots we again label these two axes as $x$ and $y$). The slices
were taken through the maxima of the two peaks, i.e.~at 
$x_1,x_4$ = 5,8 (left column of plots), respectively at $x_1,x_4$ = 15,15
(right column of plots). We use values of $\zeta = 0,0.3,0.5$
showing the two extremal situations where only one of the lumps dominates
($\zeta = 0$, top) and $\zeta = 0.5$, bottom) and an intermediate
situation ($\zeta = 0.3$, middle). Note the different scale for the 
two columns of plots. The data we show are from the run
with the boundary phase attached in 1-direction. 
The plots show clearly that the zero-mode changes its location 
with $\zeta$. The same pattern with certain values of 
$\zeta$ where a single lump dominates and intermediate values with
an advancing and a retreating lump is also seen for the
other configurations which visit two or three lumps. Equivalent plots taken
for the zero-modes computed with boundary conditions in 2-direction
show similar behavior. The lumps sit at the same positions 
and even have similar shape, such as e.g.\ the slightly elongated 
form along the $x$-axis in the plots in the left columns. 

\subsection{Results from the whole sample}

\begin{figure}[t!]
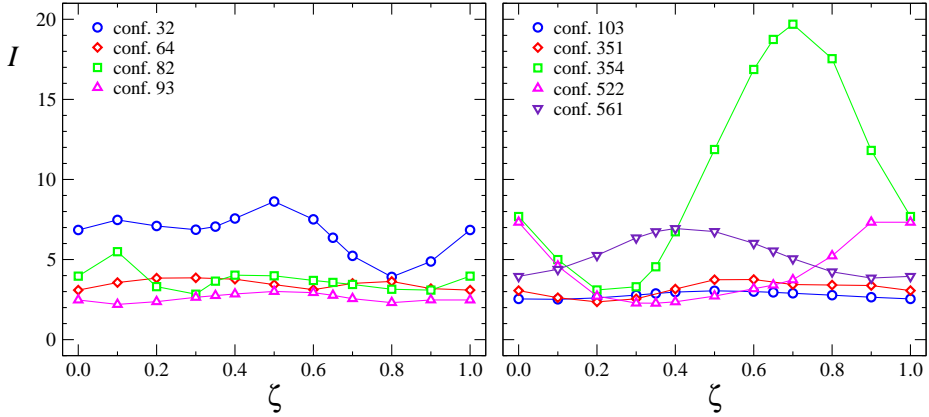

\vspace{5mm}
\begin{center}
\hspace*{-3mm}
\epsfig{file=ipr_vs_phase_z16_b845_a.eps,height=5.5cm,clip}
\hspace{-1mm}
\epsfig{file=ipr_vs_phase_z16_b845_b.eps,height=5.5cm,clip}
\end{center}
\vspace{-3mm}
\caption{Inverse participation ratio $I$ as a function of $\zeta$ for
the $16^4$, $\beta = 8.45$ ensemble. In order to avoid overcrowded 
figures we show the data for configurations 32, 64, 82 and 93 on the 
l.h.s., and configurations 103, 351, 354, 522, 561 on the r.h.s.
\label{ipr_vs_phase_z16_b845}}
\end{figure}

Let us now try to analyze the common features of all configurations
in the ensemble. We start with discussing the inverse participation 
ratio as a function of $\zeta$. In Fig.\ \ref{ipr_vs_phase_z16_b845} we
show two such figures where we again distribute the 10 configurations among
two plots, with the l.h.s.\ plot containing configurations 32, 64, 82 and 93
and configurations 103, 351, 354, 522, 561 on the r.h.s. 
We remark that configuration 
61 which is also listed in Table \ref{confsz16b845} has values of
$I$ above 50 for all $\zeta$ such that 
this defect-like lump does not show up for the plot-range we have chosen. 

The plots show that some of the configurations display quite a sizeable
change in their localization as a function of $\zeta$. Four of the zero-modes
(64, 93, 103, 351) show only relatively
small changes of their inverse participation
ratio. An inspection of Table \ref{confsz16b845} shows that this
behavior seems to be correlated with neither $d_4$ nor with the
number of lumps visited by the zero-mode.
 
\begin{figure}[t!]
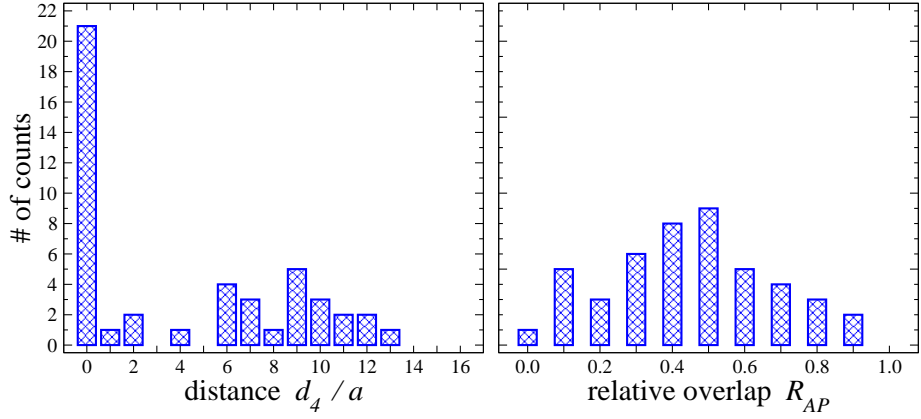

\begin{center}
\vspace{5mm}
\hspace*{-5mm}
\epsfig{file=disthistoz16b845.eps,height=5.5cm,clip}
\epsfig{file=overlaphistoz16b845.eps,height=5.5cm,clip}
\end{center}
\vspace{-3mm}
\caption{Histograms for the distance between the peaks in the scalar density 
when comparing periodic to anti-periodic b.c.\ (l.h.s.\ plot) 
and for the relative overlap (r.h.s.\ plot). The data are for the 
ensemble on the torus ($16^4$, $\beta = 8.45$).
\label{disthistoz16b845}}
\end{figure}

Let us now address the question whether changing its location is
a common property of the zero-modes for our ensemble on the torus.
We do this by again analyzing histograms for the distance $d_4$ between
the maxima of the periodic and anti-periodic scalar density and 
histograms for the overlap between the two lumps. In Fig.\
\ref{disthistoz16b845} we show such plots with the histogram for
the distance on the l.h.s.\ and the histogram for the overlap 
on the r.h.s. The histograms were taken from all 46 configurations
in our ensemble on the torus.
 
The histogram for the distance $d_4$ shows that indeed configurations
are quite abundant 
where the zero-mode changes its position by a sizeable amount,
larger than a simple deformation by a fluctuation. In particular about
half of the configurations show a $d_4$ of at least 6 in lattice units, 
i.e.~the centers of the lumps are at least 0.56 fm apart. The distance $d_4$
reaches values up to 13 in lattice units, i.e.\ 1.22 fm. The histogram 
for the relative overlap shows that even most of the zero-modes that do
not change their location suffer some deformation when changing the 
boundary condition, such that the relative overlap is pushed below 1. 

\begin{figure}[t!]
\begin{center}
\vspace{5mm}
\hspace*{-5mm}
\epsfig{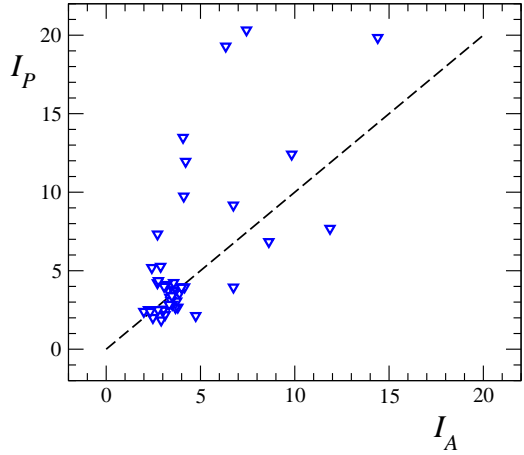}
\end{center}
\vspace{-3mm}
\caption{Scatter plot for the inverse participation ratio of the
zero mode with periodic and anti-periodic boundary condition
($I_P$ respectively $I_A$). 
The data (triangles) are for the ensemble on the torus ($16^4$, $\beta = 8.45$).
We display the line with $I_P = I_A$ as a dashed line.   
\label{iprscatterz16b845}}
\end{figure}
Finally in Fig.\ \ref{iprscatterz16b845} we show a scatter plot 
of the periodic inverse participation ratio $I_P$ versus its
anti-periodic counterpart $I_A$. For the deconfined phase the equivalent
plot shown in Fig.\ \ref{iprscatter} has revealed an asymmetry 
between periodic and anti-periodic boundary conditions which 
is closely related to the critical values of $\zeta$ where the zero-mode
can change its location.
For the ensemble on the torus our
Fig.\ \ref{iprscatterz16b845} does not reveal such a structure
and the data essentially scatter symmetrically around the line 
$I_P = I_A$ which we display as a dashed straight line. This indicates
that for the torus we do not find a pronounced pattern which links the 
localization of the mode to particular values of $\zeta$.

\section{Summary}

In this article we have studied properties of topological excitations
of SU(3) lattice gauge configurations with topological charge $Q = \pm 1$.
Through the index theorem such configurations give rise to a 
single zero-mode of the Dirac operator. The zero-mode is localized 
on the topological excitation and reflects the properties of the 
underlying lump in the gauge field. We analyzed how the zero-mode changes when 
applying an arbitrary phase $\exp(i 2\pi \zeta)$ at the temporal boundary 
condition for the Dirac operator. Our main findings are:

\begin{itemize}

\item For the ensemble in the deconfined phase we find that the zero-mode 
can change its position at $\zeta = 0$ for configurations with 
real Polyakov loop and at $\zeta = 1/3, \zeta = 2/3$ for configurations
in the complex sector.

\item The zero-mode is very spread out at these critical values 
and most localized for $\zeta = 0.5$ for real Polyakov loop, 
respectively $\zeta = 5/6, \zeta = 1/6$ for the complex sectors.

\item These observations nicely match the predictions for zero-modes
of KvB solutions. 

\item An analysis of the spectral gap shows the same periodic
behavior in $\zeta$ supporting earlier findings of only a single 
transition temperature for chiral symmetry restoration in all sectors
of the Polyakov loop.    

\item For the ensemble in the confined phase 
just below $T_c$ we find that for
a large portion of our ensemble the zero-mode is located at different
positions when changing the boundary condition. 

\item Also the localization of the mode fluctuates considerably during 
a complete cycle through $\zeta$.

\item When analyzing the time dependence one finds that the zero-modes are 
stretched along the time direction for both ensembles above and 
below $T_c$.

\item Several properties of zero-modes in the confined phase match 
predictions for
KvB zero modes, but the resemblance is not as close as in the deconfined 
phase.

\item Also the zero-modes for configurations on the torus change their location
when changing the phase at the boundary. 

\item The results obtained when applying the boundary condition at different
directions are consistent with each other.

\item About half of our torus configurations have zero-modes
located at different positions
when comparing periodic to anti-periodic boundary conditions.

\end{itemize}

Our results show that for a large portion of our 
configurations an excitation with topological 
charge $|Q|$ = 1 is not a single lump. 
Instead it is built from several independent objects and does not 
resemble a simply connected instanton-like object.
\\
\\
{\bf Acknowldegements:} \\
We thank Meinulf G\"ockeler, Michael M\"uller-Preussker,
Paul Rakow, Andreas Sch\"afer and Christian Weiss for discussions.
We are particularly greatful for the ongoing interest from  
Pierre van Baal who generously shared his insight and experience
with us. Christof Gattringer thanks the 
Austrian Academy of Science for support 
(APART 654). This project was partly supported by 
BMBF and DFG. The calculations were done on the Hitachi SR8000
at the Leibniz Rechenzentrum in Munich and we thank the LRZ staff for
training and support.

\end{document}